\newcommand{\kaonangle}{$\cos\theta_\mathrm{CM}^{K}$}
\documentclass[epj]{svjour}
\usepackage{arydshln}
\usepackage{graphics}
\usepackage{graphicx}
\usepackage{xcolor}
\usepackage{sidecap}
\usepackage{longtable}
\usepackage{lineno}
\begin{document}
\title{$K^+\Lambda$ photoproduction at forward angles and low momentum transfer}

\author{%
	 S.~Alef\inst{1}%
	\and P.~Bauer\inst{1}%
	\and D.~Bayadilov\inst{2,3}%
	\and R.~Beck\inst{2}%
	\and A.~Bella\inst{1,}\thanks{No longer employed in academia}%
	\and J.~Bieling\inst{2,\mathrm{a}}%
	\and A.~Braghieri\inst{4}%
	\and P.L.~Cole\inst{5}
	\and{D.~Elsner}\inst{1}
	\and \newline R.~Di Salvo\inst{6}%
	\and A.~Fantini\inst{6,7}%
	\and O.~Freyermuth\inst{1}%
	\and F.~Frommberger\inst{1}%
	\and F.~Ghio\inst{8,9}%
	\and S.~Goertz\inst{1}%
	\and A.~Gridnev\inst{3}%
	\and D.~Hammann\inst{1,\mathrm{a}}%
	\and J.~Hannappel \inst{1,}\thanks{Currently, DESY Research Centre, Hamburg, Germany}%
	 \and T.C.~Jude\inst{1,}\thanks{Corresponding author:  jude@physik.uni-bonn.de}%
	\and K.~Kohl\inst{1}
	\and N.~Kozlenko\inst{3}
	\and A.~Lapik\inst{10}
	\and P.~Levi Sandri\inst{11}%
	\and V.~Lisin\inst{10}%
	\and G.~Mandaglio\inst{12,13}%
	\and F.~Messi\inst{1,\mathrm{a}}%
	\and R.~Messi\inst{6,7}%
	\and D.~Moricciani\inst{11}%
	\and V.~Nedorezov\inst{10}%
	\and V.A~Nikonov\inst{2,3,}\thanks{Deceased}%
	\and D.~Novinskiy\inst{3}%
	\and P.~Pedroni\inst{4}%
	\and A.~Polonskiy\inst{10}
	\and B.-E.~Reitz\inst{1,\mathrm{a}}%
	\and M.~Romaniuk\inst{6,14}
	\and A.V~Sarantsev\inst{2,3}
	\and G.~Scheluchin\inst{1}%
	\and H.~Schmieden\inst{1}%
	\and A.~Stuglev\inst{3}%
	\and V.~Sumachev\inst{3,\mathrm{d}}
	\and V.~Vegna\inst{1,a}%
	\and V.~Tarakanov\inst{3}
	\and T.~Zimmermann\inst{1,\mathrm{a}}
}
\institute{%
	Rheinische Friedrich-Wilhelms-Universit\"at Bonn, Physikalisches Institut, Nu\ss allee 12, 53115 Bonn, Germany
	\and Rheinische Friedrich-Wilhelms-Universit\"at Bonn, Helmholtz-Institut f\"ur Strahlen- und Kernphysik, Nu\ss allee 14-16, 53115 Bonn, Germany%
		\and Petersburg Nuclear Physics Institute, Gatchina, Leningrad District, 188300, Russia
	\and INFN sezione di Pavia, Via Agostino Bassi, 6 - 27100 Pavia, Italy
	\and Lamar University, Department of Physics, Beaumont, Texas, 77710, USA
	\and INFN Roma ``Tor Vergata", Via della Ricerca Scientifica 1, 00133, Rome, Italy
	\and Universit\`a di Roma ``Tor Vergata'', Dipartimento di Fisica, Via della Ricerca Scientifica 1, 00133, Rome, Italy
	\and INFN sezione di Roma La Sapienza, P.le Aldo Moro 2, 00185, Rome, Italy 
	\and Istituto Superiore di Sanit\`a, Viale Regina Elena 299, 00161, Rome, Italy 
		\and Russian Academy of Sciences Institute for Nuclear Research, Prospekt 60-letiya Oktyabrya 7a, 117312, Moscow, Russia
	\and INFN - Laboratori Nazionali di Frascati, Via E. Fermi 54, 00044, Frascati, Italy
	\and INFN sezione Catania, 95129, Catania, Italy
	\and Universit\`a degli Studi di Messina, Dipartimento MIFT,  Via F. S. D'Alcontres 31, 98166, Messina, Italy
	\and Institute for Nuclear Research of NASU, 03028, Kyiv, Ukraine
}
\date{Received: date / Revised version: date}
%
\abstract{
	$\gamma p \rightarrow K^+ \Lambda$ differential cross sections and recoil polarisation data from threshold for extremely forward angles are presented.  The measurements were performed at the BGOOD experiment at ELSA, utilising the high angular and momentum resolution forward spectrometer for charged particle identification.
	The high statistics and forward angle acceptance enables the extraction of the cross section as the minimum momentum transfer to the recoiling hyperon is approached.
}
\maketitle

\section{Introduction}

Associated strangeness  ($KY$) photoproduction is a crucial area of study to elucidate the nucleon excitation spectrum and the relevant degrees of freedom. 
There remain many resonances predicted by constituent quark models (CQMs)~\cite{capstick86,capstick92,capstick94,riska01}, lattice QCD calculations~\cite{edwards11}, harmonic oscillator and hypercentral CQMs~\cite{klempt12,giannini15} and Dyson-Schwinger equations of QCD~\cite{roberts11} that have not been observed experimentally.  Significant advancements however have been made, both in the understanding of known resonances properties and new resonance discoveries\footnote{The Particle Data Group, for example,  recognised 10 \textit{four star} and 3 \textit{three star} $N^*$ resonances above ground state in 2010, compared to 13 and 7 in 2020~\cite{pdg2010,pdg20}.}.
A main motivation of the study of $KY$ photoproduction channels over the last 15 years has been to search for these ``missing resonances” which may only couple weakly to $N\pi$ final states~\cite{capstick00,loering01}. 
The ensuring wealth of high statistics data from the Crystal Ball @ MAMI~\cite{jude14}, CLAS~\cite{bradford06,mccracken10,dey10,bradford07,mcnabb04,carman09}, SAPHIR~\cite{glander04}, LEPS~\cite{sumihama06,shiu18} and GRAAL~\cite{lleres07} 
collaborations have rendered the $KY$ channels the closest to a ``complete experiment”, 
where a judiciously selected set of polarisation observables permit a complete description of the photoproduction mechanism~\cite{barker75}.  This is partly due to the weak, self analysing decay of the $\Lambda$ enabling easier access to the recoiling baryon (single and double) polarisation observables.  Despite this data and support from partial wave analyses (PWA) with dynamical coupled-channel frame works~\cite{anisovich07,anisovich14,muller19,roenchen18}, isobar models~\cite{skoupil16,skoupil18,mart99,clymton17,lee01,janssen01,janssen01EPJA,janssen03}, and models incorporating Regge trajectories~\cite{cruz12a,cruz12b,bydovsky19} to fix $t$-channel contributions using data above the resonance region (photon beam energies larger than 4\,GeV), a mutually consistent description between theory and data of $KY$ photoproduction channels has not been realised.

The $K^+\Lambda$ threshold at a centre of mass energy  of 1609\,MeV, is in the third resonance region where
an abundance of $s$-channel resonances up to high spin states, $u$-channel hyperon resonances and $t$-channel $K$, $K^*$ and $K_1$ exchanges contribute.  The isospin singlet $\Lambda$, however, acts as a  filter to remove intermediate $\Delta^*$ states which are present in $K\Sigma$ channels, enabling a ``cleaner" study of $t$-channel processes.
At forward angles, where the cosine of the centre of mass $K^+$ polar angle, \kaonangle, exceeds 0.9, there is a paucity of data to constrain the reaction mechanism, and the existing cross section data of SAPHIR~\cite{glander04} and CLAS~\cite{bradford06,mccracken10,mcnabb04}  have  pronounced inconsistencies\footnote{The LEPS collaboration data~\cite{sumihama06,shiu18} starts at a photon beam energy of 1.5\,GeV and is generally in agreement with CLAS data.}.  
This has led to a poor understanding of the dynamics of the Born terms and $t$-channel $K^+$ and $K^*$ exchanges which dominate at forward angles (see for example ref.~\cite{bydzovsky12}).
PWA solutions have  also included different $s$-channel resonance contributions, depending if the fits used the SAPHIR or CLAS datasets (see for example ref.~\cite{mart06}).  Data with high \kaonangle{} resolution at forward (and backward) angles is also sensitive to high-spin intermediate states, where the corresponding Legendre polynomials change quickly with respect to \kaonangle.  States with spin 5/2 and 7/2  have been incorporated in previous PWA and isobar model solutions  (see for example refs.~\cite{anisovich07,anisovich14,mart06}). 



Forward angle kinematics also enables access to a regime where the momentum transfer to the recoiling hyperon is minimised.  This is a vital input for the description of hypernuclei electroproduction at low $Q^2$~\cite{achenbach12,achenbach12b,garibaldi19,motoba10,bydovsky07,bydzovsky12hypernuclei}.  Studying the $Y$-$N$ interaction is crucial for an SU(3)$_\mathrm{flavour}$ description of baryon interactions
and provides important astrophysical constraints, for example upon the equation of state for neutron stars (see ref.~\cite{haidenbauer17} and references therein).

\begin{figure} [htp]
	\centering
	\vspace*{0cm}
	\resizebox{\columnwidth}{!}{%
		\includegraphics{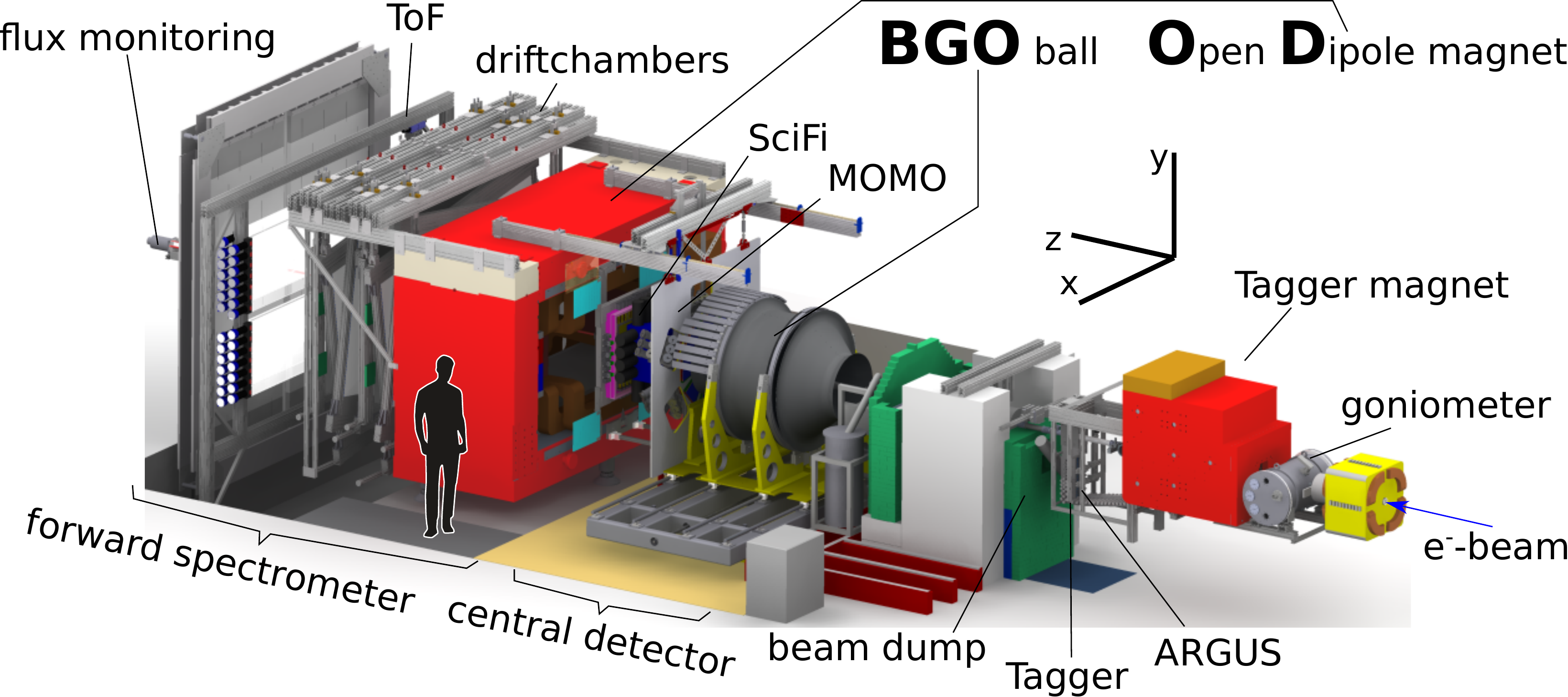}
	}
	\caption{Overview of the BGOOD setup.  The central detector region consists of the BGO Rugby Ball, enclosing the MWPCs, Plastic Scintillating Barrel and the target.  Figure taken from ref.~\cite{technicalpaper}.}
	\label{fig:BGOODsetup}
\end{figure}

The BGOOD experiment~\cite{technicalpaper} (shown in fig.~\ref{fig:BGOODsetup}) at the ELSA facility~\cite{hillert06,hillert17}
in Bonn, Germany, is ideally suited for $\gamma p \rightarrow K^+\Lambda$ measurements at forward angles.  BGOOD is composed of two distinct parts:  a forward magnetic spectrometer, ideal for the detection of forward going $K^+$, and a central calorimeter, suited for the identification of hyperons at low momentum, decaying almost isotropically.
The presented data resolve discrepancies in existing  datasets
for \kaonangle$ > 0.9$ from threshold to a centre of mass energy, $W = 1870$\,MeV.
Due to the high \kaonangle{}
resolution, the cross section as the minimum momentum transfer is approached can be determined in 0.02 \kaonangle{} intervals. 

This paper is organised as follows: sect.~\ref{sec:detector} describes the BGOOD experiment and the running conditions during the data taking. Section~\ref{sec:selectevents} explains the identification of the reaction channel and corresponding systematic uncertainties. Differential cross sections and recoil polarisation measurements are presented and discussed in sect.~\ref{sec:results}. Concluding remarks are made in sect.~\ref{sec:conclusions}.

\section{BGOOD setup and experimental running conditions}\label{sec:detector}


A detailed description of the experimental setup, performance and analysis procedures is given in ref.~\cite{technicalpaper}.

The data were taken during a 22 day beam time, using an incident ELSA electron beam energy of 3.2\,GeV and a 6\,cm long liquid hydrogen target. The electron beam was incident upon a thin crystal radiator to produce a continuous spectrum of bremsstrahlung photons.  The orientation of the crystal was such that a coherent, polarised peak was set at a photon beam energy ($E_\gamma$) of 1440\,MeV, however the polarisation was not required for the presented analysis. 
The energy of each photon was determined by momentum analysing the post-bremsstrahlung electron in the \textit{Photon Tagger}.  This consists of a dipole magnet and a hodoscope of plastic scintillators to detect the deflection angle of the electron.  Photon energies were measured from 10\,\% to 90\,\% of the extracted ELSA electron beam energy.

The photon beam passed through a 7\,mm diameter collimator, with approximately 80\,\% of the bremsstrahlung photons impinging upon the target (referred to as the \textit{tagging efficiency}). 
The photon flux was determined continually during the data taking using the \textit{Flumo} detector downstream from the experiment. 
This consists of two sets of three plastic scintillators arranged downstream from each other to detect electron-positrons from pair production in the beam. Flumo was calibrated to the photon flux by taking separate, low rate runs using a lead glass scintillator, \textit{GIM}, with 100\,\% photon detection efficiency. 
The integrated photon flux from  900 to 1500\,MeV photon beam energy (the approximate region of the data shown) was $8.4\times10^{12}$.


The \textit{BGO Rugby Ball}, comprised of 480 BGO crystals individually coupled to photomultipliers, covers polar angles 25$^\circ$ to 155$^\circ$.
The fast time read out per crystal allows clean identification of neutral meson decays to photons.

A set of two coaxial and cylindrical multiwire proportional chambers (\textit{MWPCs}) and a
\textit{Plastic Scintillating Barrel} surround the target within the BGO Rugby Ball
and are used for charged particle identification and reaction vertex reconstruction.

The \textit{Forward Spectrometer} is a combination of tracking detectors, an open dipole magnet and time of flight walls.  Two scintillating fibre detectors, \textit{MOMO} and \textit{SciFi}, track particles from the reaction vertex in the target. 
Downstream from these is the \textit{Open Dipole Magnet}, operating at an integrated field strength of 0.7\,Tm and covering polar angles 1$^\circ$ to 12$^\circ$ or 8$^\circ$ in the horizontal or vertical planes respectively. 
Particle trajectories downstream from the Open Dipole Magnet are determined using eight double layered drift chambers, and  
 particle momentum is subsequently determined by the deflection of the trajectory in the magnetic field. 
Three time of flight (\textit{ToF}) walls at the end of the spectrometer measure particle $\beta$. 

The region between the BGO Rugby Ball and the Forward Spectrometer is covered by the \textit{SciRi} detector, which is composed of three segmented rings of plastic scintillators for charged particle detection.  SciRi covers a polar angle range of 10$^\circ$ to 25$^\circ$.

\section{Event selection}\label{sec:selectevents}
$K^+$  were identified in the Forward Spectrometer from spatial coincidences between MOMO, SciFi, the Drift Chambers and the ToF
walls. The momentum calculation used a three dimensional magnetic field description, including fringe fields extending beyond the magnet yoke, and particle energy loss from the target, air and detector materials. 
The particle trajectory was ``stepped through” in discrete intervals, applying the expected acceleration due to the Lorentz force and material energy loss.  The interval lengths were dynamically determined to optimise accuracy and computational time depending upon the magnitude of the energy loss and Lorentz force per interval.  An iterative approach was used to determine the optimum trajectory and momentum, given the hit positions in the detectors and weighted by their spatial resolutions.  A momentum resolution of approximately 5\,\% of the measured momentum was achieved.  See ref.~\cite{technicalpaper} for details.

Particle $\beta$ was determined by time measurements in the ToF walls, accounting for the trajectory length and particle energy loss. Contrary to the default track finding routine described in ref.~\cite{technicalpaper}, a cluster in MOMO was not required to form a forward track  due to an efficiency of only 80\,\%.  If no MOMO cluster was identified, it was sufficient to use only a SciFi cluster and the target centre as a space point.  The increase in background and reduction in spatial resolution were proved to be negligible. 

\begin{figure} [h]
	\centering
		\resizebox{0.8\columnwidth}{!}{
		\includegraphics{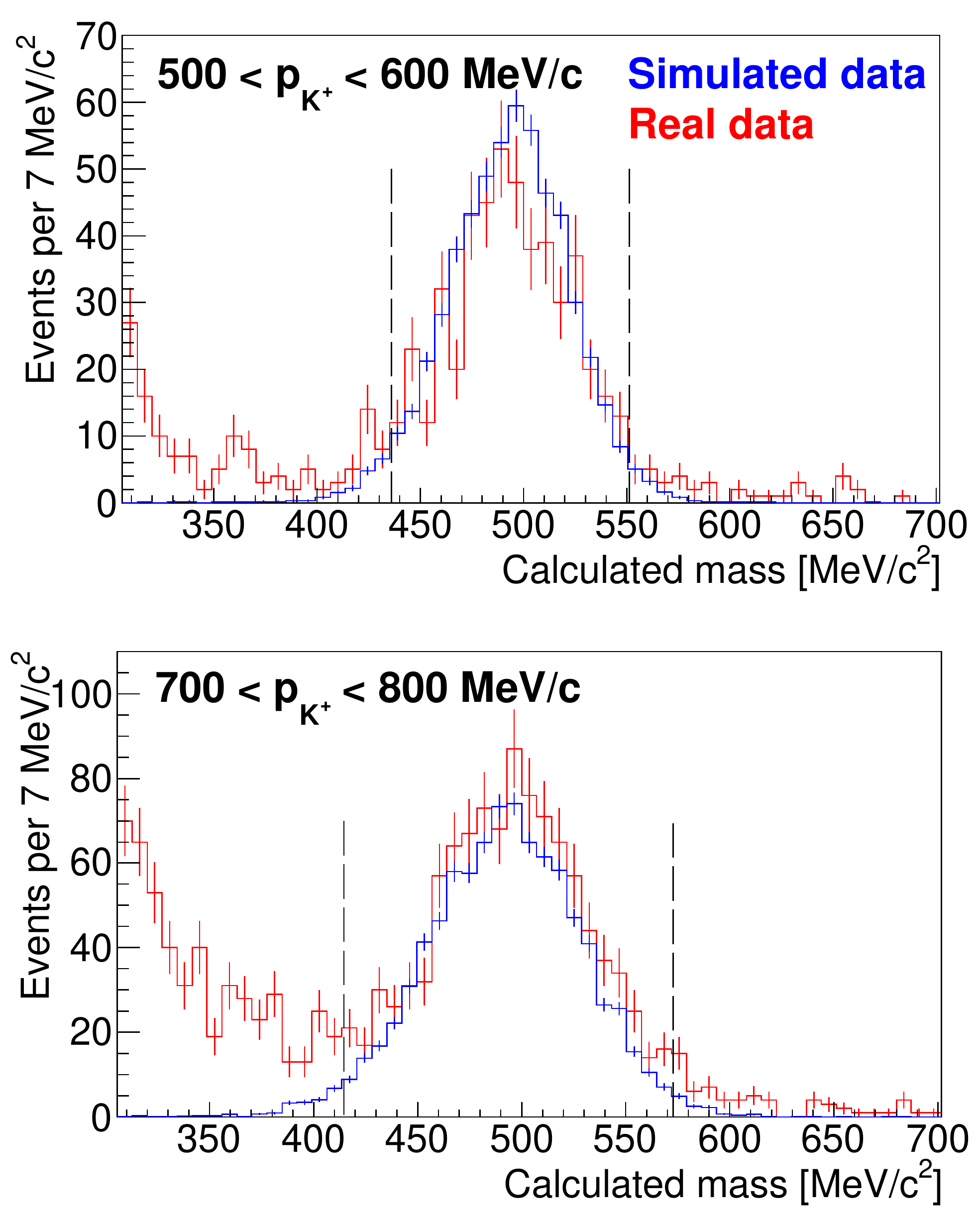} 
	}
	\caption{Mass reconstruction for $K^+$ candidates in the forward spectrometer for real and simulated data (red and blue lines respectively).  The $K^+$ momentum, $p_{K^+}$, intervals are labelled inset.  The dashed lines indicate the selection cut for the median value of $p_{K^+}$ described in the text.}
	\label{fig:massselection}
\end{figure}

The mass of forward particles was calculated from momentum and $\beta$. 
Figure~\ref{fig:massselection} shows two examples of the reconstructed $K^+$ mass for different momentum intervals, with good agreement between real and simulated events.  The rising structure towards low masses at 300\,MeV/c$^2$ in the real data
is from $\pi^+$ from other hadronic reactions, and positrons from pair production in the beam.  The small peak at 360\,MeV/c$^2$ in the lower momentum interval is from pair production in the beam from an ELSA electron bunch adjacent in time (every 2\,ns) to the bunch containing the electron responsible for the triggered event.  Timing cuts with respect to particle $\beta$ remove most of these events, however these selection cuts are very conservative with respect to detector time resolutions to avoid removing any particles from triggered hadronic reactions.

Candidate events were selected over $\pm2\sigma$ of the reconstructed $K^+$ mass by approximately fitting a Gaussian function to the mass distribution.
This varied with $K^+$ momentum, from $\pm 47$\,MeV/c$^2$ 
and $\pm 106$\,MeV/c$^2$ at 450\,MeV/c and 1000\,MeV/c respectively.

\begin{figure} [h]
	\centering
	\vspace*{0cm}
	\resizebox{0.8\columnwidth}{!}{%
		\includegraphics{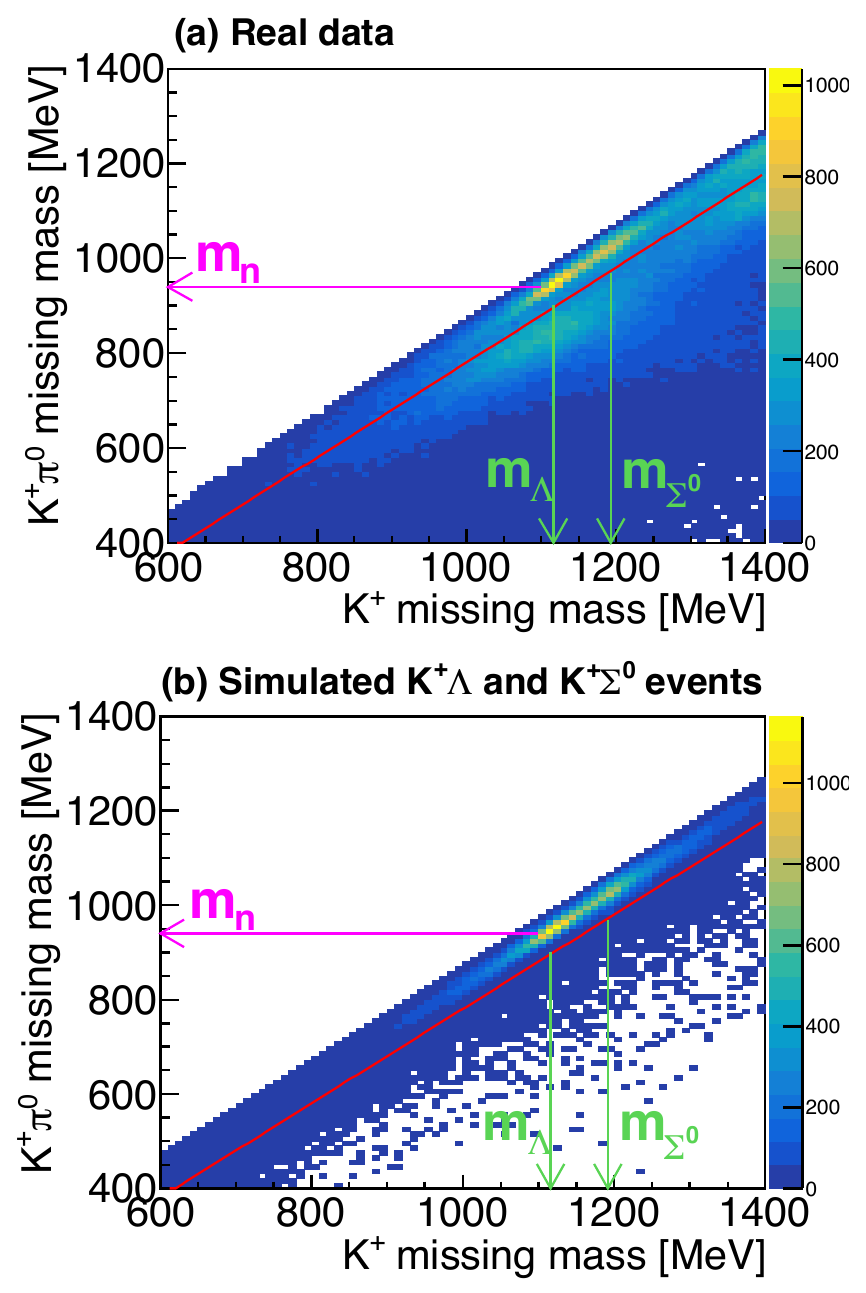} 
	}
	\caption{Missing mass recoiling from the $K^+\pi^0$ system versus the missing mass from the $K^+$. (a) Real data. (b) Simulated $K^+\Lambda$ and $K^+\Sigma^0$ events, approximately weighted to the measured ratio.
	 Events were selected above the red line.
	}
	\label{fig:pionselection}
\end{figure}

Due the relatively small cross section compared to non-strange channels, identification of the decay $\Lambda\rightarrow \pi^0 n$  was required to enhance the signal relative to background.
$\pi^0$ were identified in the BGO Rugby Ball via the two photon decay, where the measured invariant mass was required to be $\pm 30$\,MeV/c$^2$ from the accepted $\pi^0$ mass, corresponding to $\pm2\sigma$.  Figure~\ref{fig:pionselection} shows the  missing mass from the $K^+\pi^0$ system corresponding to the neutron mass for the $K^+\Lambda$ channel, plotted against the missing mass from the forward $K^+$. Events were selected above the red line.

\begin{figure} [h]
	\centering
	\vspace*{0cm}
	\resizebox{\columnwidth}{!}{%
		\includegraphics{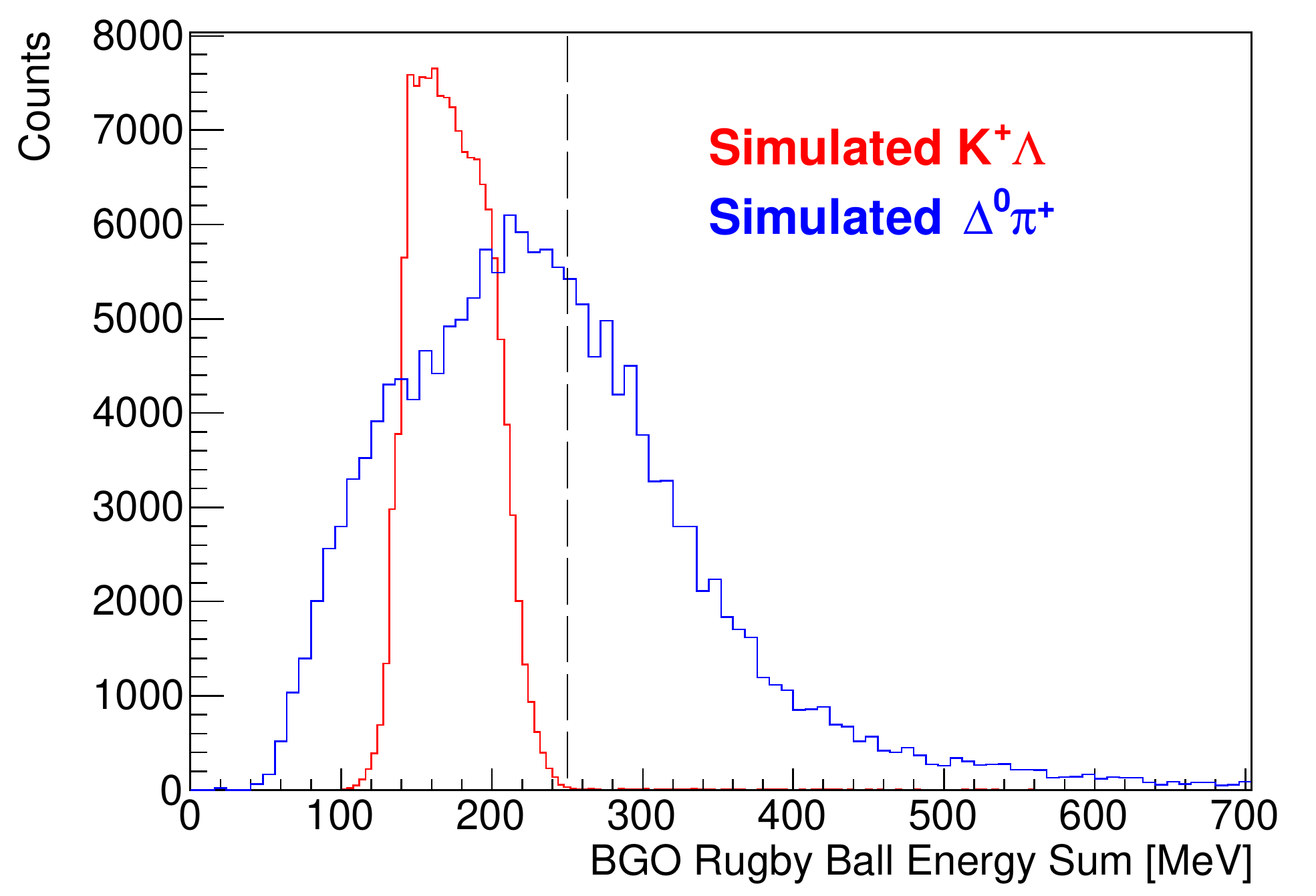} 
	}
	\caption{Total energy deposition in the BGO Rugby Ball for simulated $\gamma p \rightarrow K^+\Lambda$ and $\gamma p \rightarrow \Delta^0\pi^+$ events (red and blue lines respectively) when a $K^+$ candidate was identified in the forward spectrometer and the $\pi^0$ from the $\Lambda$ decay in the BGO Rugby Ball.  The dashed black line indicates the maximum energy deposition allowed when selecting $K^+\Lambda$ events.
	}
	\label{fig:esum}
\end{figure}

 Events were rejected if a charged particle was identified in either the BGO Rugby Ball (via coincidence with the plastic scintillating barrel) or the intermediate SciRi detector.  The total energy deposition in the BGO Rugby Ball was also required to be lower than 250\,MeV.
 The simulated data shown in fig.~\ref{fig:esum} demonstrates this removes approximately half of
 the most significant background from falsely identified $\pi^+$ from $\Delta^0\pi^+$ events.

\begin{figure} [h]
	\centering
	\vspace*{0cm}
		\resizebox{\columnwidth}{!}{%
\includegraphics{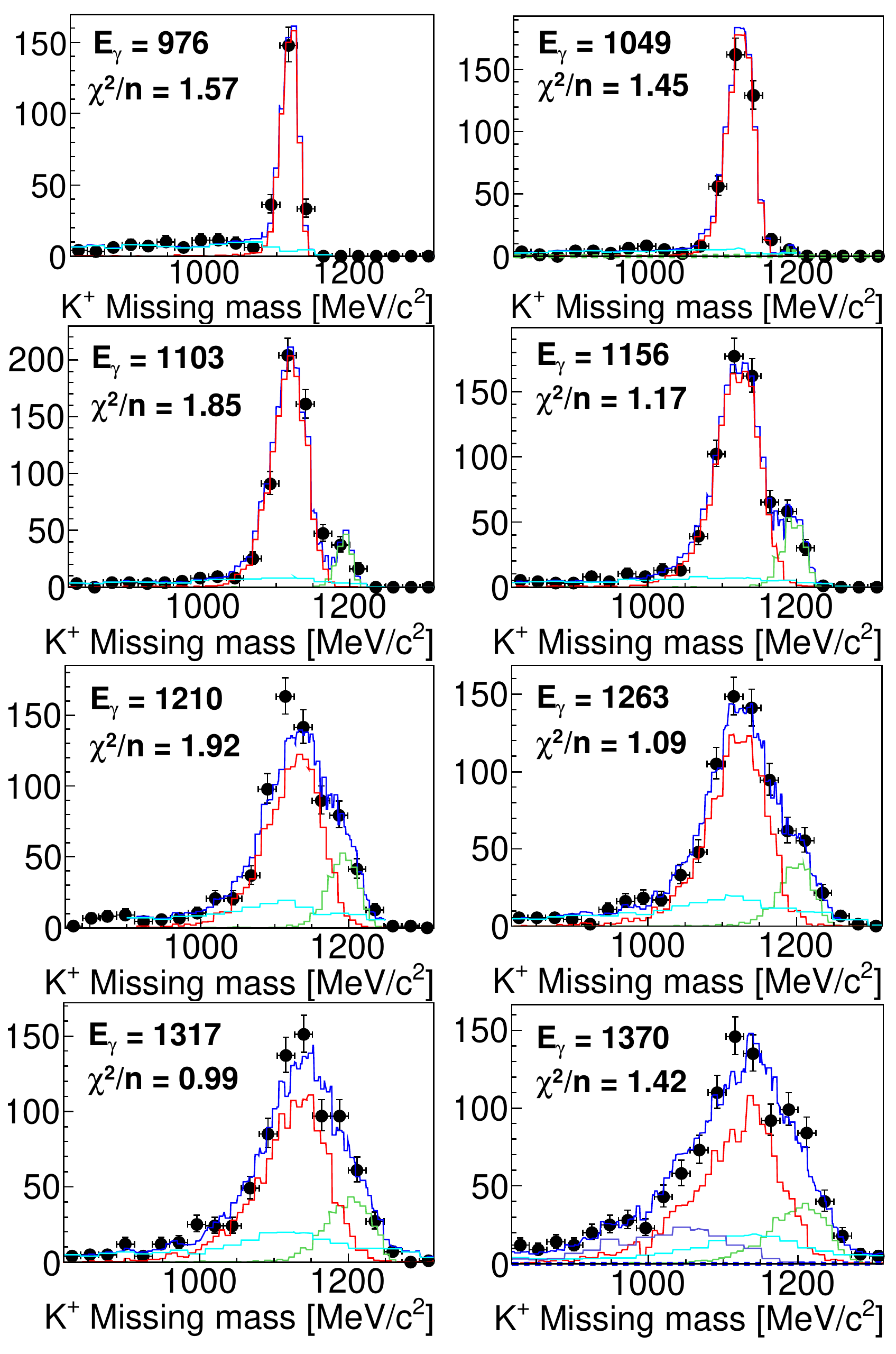} 
}
	\caption{Missing mass from forward $K^+$ candidates after selection criteria described in the text.  Every other photon beam energy bin ($E_\gamma$) is shown and labelled in units of MeV, with corresponding reduced $\chi^2$ for the fit.   The data are the black points, with fitted spectra from simulated $K^+\Lambda$ and $K^+\Sigma^0$ and $e^+$/$\pi^+$ background (red, green and cyan lines respectively).  The blue line is the summed total fit.  The highest energy bin, $E_\gamma = 1370$\,MeV also includes the simulated $\Delta^0\pi^+$ contribution (purple line).}
	\label{fig:missingmass}
\end{figure}

Figure~\ref{fig:missingmass} shows the $K^+$ missing mass for different photon beam intervals. The distribution of the $\pi^+$ and $e^+$ background was described by an equivalent analysis of negatively charged particles, where $\pi^-$ and $e^-$ have similar kinematics.  Simulated data were used to describe the $K^+\Lambda$ signal and the $K^+\Sigma^0$ background.
The simulations followed energy and angular distributions from previously measured cross sections~\cite{mccracken10,dey10}, however the intervals in \kaonangle{} and energy were sufficiently small so that the missing mass spectra could be considered fixed across each interval.
The spectra therefore depended solely on the experimental energy and spatial resolutions, and accurately described the real data.
A fit was subsequently applied using the three missing mass spectra as templates with separate scaling factors in order to extract the $K^+\Lambda$ yield.

\begin{figure} [h]
	\centering
	\vspace*{0cm}
	\resizebox{0.9\columnwidth}{!}{%
		\includegraphics{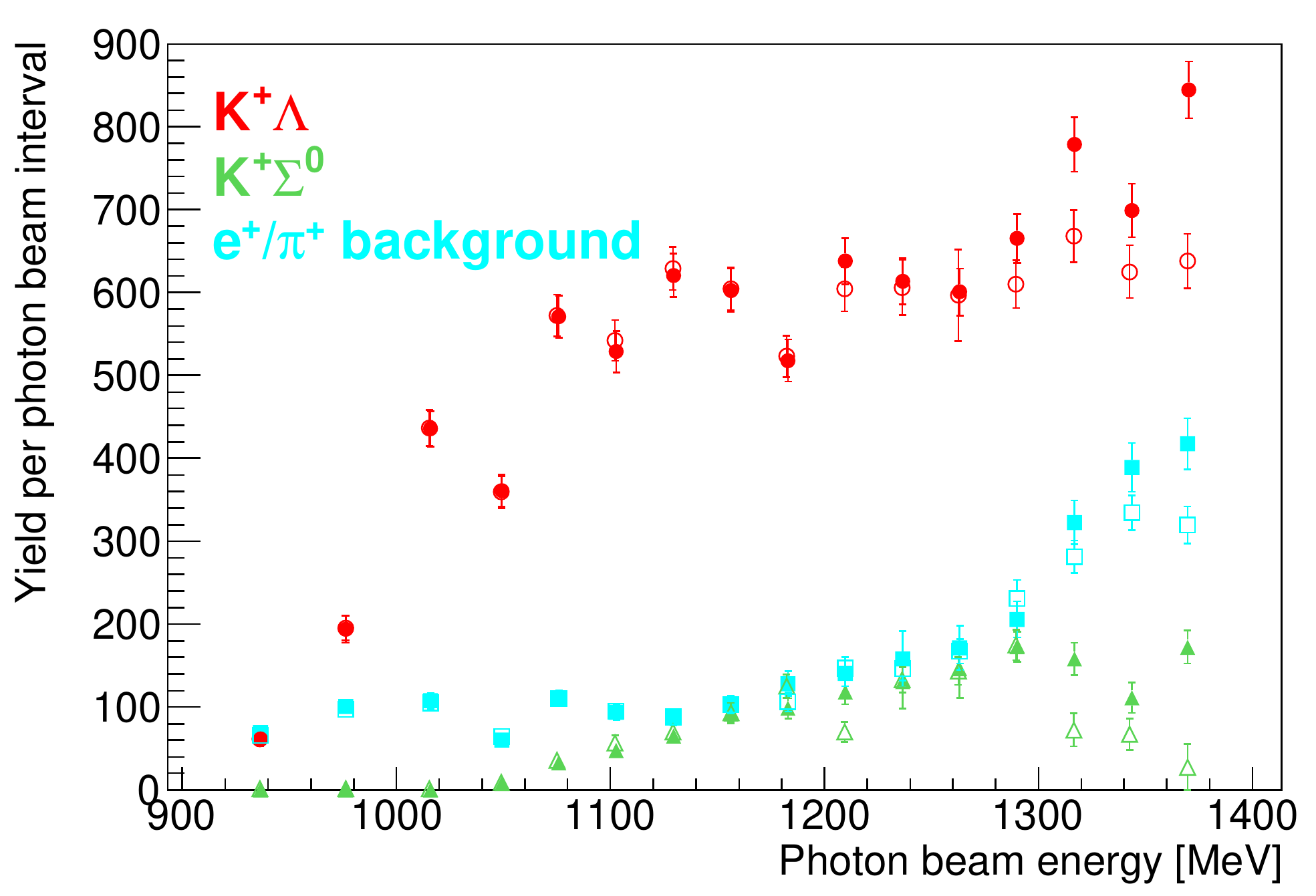} 
	}
	\caption{The extracted yields for the $K^+\Lambda$ signal and background from $K^+\Sigma^0$ and $e^+\pi^+$ misidentification (red circles, green triangles  and cyan squares respectively).  The solid filled data points are without the simulated $\Delta^0\pi^+$ background, the open data points are when including this additional background.}
	\label{fig:compareyields}
\end{figure}

To fully understand background contributions, missing mass spectra from additional simulated channels were included in the fit.  The only significantly contributing channel proved to be $\gamma p \rightarrow \Delta^0\pi^+$, where the $\pi^+$ was mistaken for a $K^+$.  This was already included in the $e^+/\pi^+$ background (the cyan line in fig.~\ref{fig:missingmass}), however the inclusion of this simulated channel allowed the relative contributions of misidentified $e^+$ and $\pi^+$ to vary.
This channel only contributed in the highest four energy intervals, and did not significantly change the extracted $K^+\Lambda$ yield.  For these intervals, the fit including the additional $\Delta^0\pi^+$ missing mass spectrum was used for the $K^+\Lambda$ yield extraction if the reduced $\chi^2$ of the fit was improved.  This occurred for the highest two data points, where the reduced $\chi^2$ were 2.47 and 2.50 without including the $\Delta^0\pi^+$ spectra, and 1.45 and 1.42 when including it.
Fig.~\ref{fig:compareyields} shows the extracted yields with and without the simulated  $\Delta^0\pi^+$ data.

\subsection{Detection efficiency calculations}

The detection efficiency was determined using a Geant4~\cite{geant4}
simulation of the experimental setup. This included all spatial, energy and time resolutions, efficiencies for all detectors in the forward spectrometer (described in ref.~\cite{technicalpaper}) and the modelling of the hardware triggers described below.

Three hardware trigger conditions, listed in table~\ref{table:triggers} were implemented for a broad range of experimental requirements. Trigger 4 was used for this analysis, where approximately 80\,MeV minimum energy deposition was required in the BGO Rugby Ball and a signal in the SciFi and ToF detectors, described in table 1 as a \textit{Forward Track}.

\begin{table}[h]
	\begin{tabular}{c l}
		\hline\hline
		Trigger & Description \\
		\hline
		0 & High BGO energy sum ($\sim 200$\,MeV) \\
		1 & Low BGO energy sum ($\sim 80$\,MeV) \& SciRi\\
		3 & SciRi \& Forward Track\\
		4 & Low BGO energy sum \& Forward Track\\
		\hline\hline
	\end{tabular}
	\caption{BGOOD hardware triggers.  Each trigger also required a cluster in the Photon Tagger.  Trigger 2 is obsolete.}
	\label{table:triggers}
\end{table}

\begin{figure} [h]
	\centering
	\vspace*{0cm}
	\resizebox{0.9\columnwidth}{!}{%
		\includegraphics{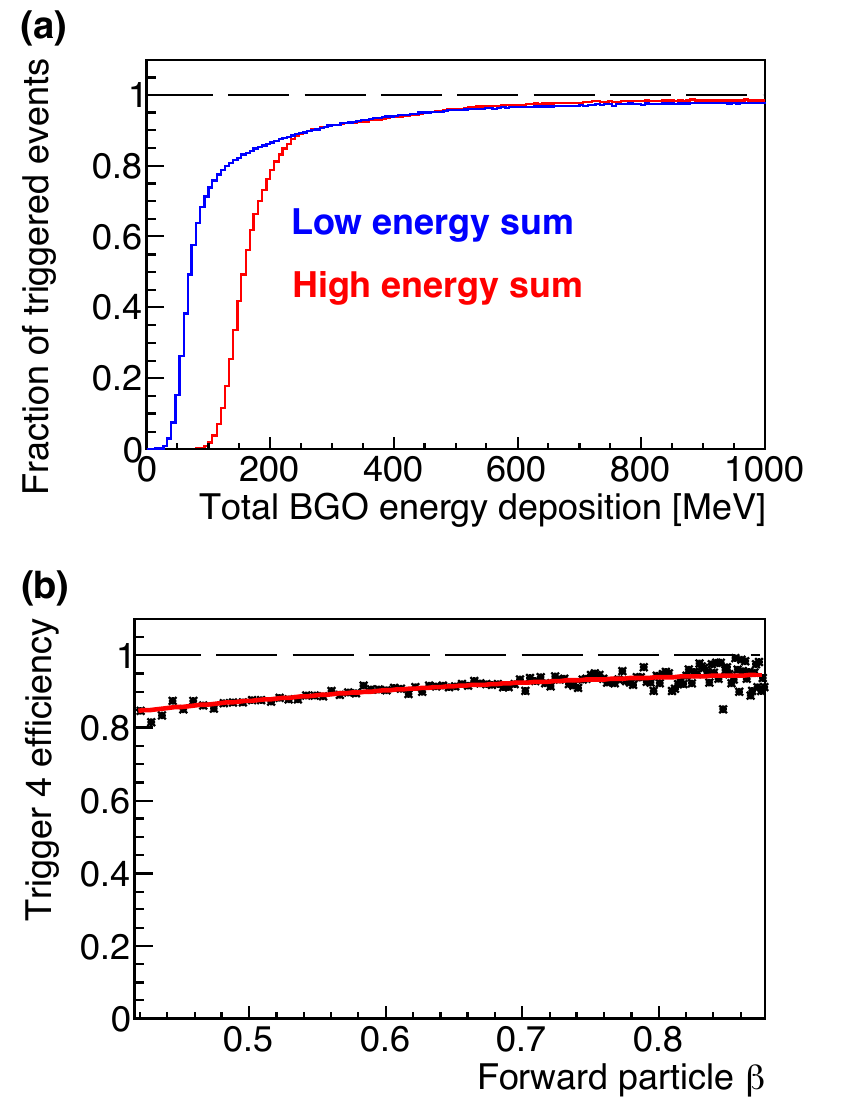} 
	}
	\caption{Modelling of the hardware triggers. (a) The fraction of events passing the low and high BGO energy sum triggers (blue and red respectively). (b) The efficiency of trigger 4 as a function of the forward going particle $\beta$.}
	\label{fig:triggereff}
\end{figure}

The efficiencies of the BGO Rugby Ball energy sum triggers, shown in fig.~\ref{fig:triggereff}(a) were determined via a ratio of events passing different trigger combinations. The high energy sum distribution was determined from the ratio of all events passing both triggers 0 and 3, and all events passing trigger 3. The low energy sum used in this analysis was determined from the ratio of all events passing both triggers 1 and 4, and all events passing trigger 3.  This ensured that the difference was dependent only upon the low energy sum efficiency, and not reaction and topologically specific.
These distributions were implemented in simulated data for an accurate determination of detection efficiencies.

Due to the small misalignment of trigger timing windows and the large time range for forward going particles, the efficiency of trigger 4 also had a small dependence upon the particle $\beta$. Fig.~\ref{fig:triggereff}(b) shows this efficiency, determined from a clean selection of forward going protons.   For forward $K^+$ from $K^+\Lambda$, $\beta$ is approximately 0.65 and 0.90 at $W = 1680$ and 1900\,MeV, corresponding to correction factors of 1.09 and 1.06 to the event yields respectively.

Both the trigger efficiency as a function of the BGO energy deposition and the $\beta$ of forward going particles were successful in describing the well known $\gamma p \rightarrow \eta p$ differential cross section, the results of which are presented in ref.~\cite{technicalpaper}.

\begin{figure} [h]
	\centering
	\vspace*{0cm}
	\resizebox{\columnwidth}{!}{%
		\includegraphics[width=\columnwidth,trim={0cm 0cm 1.5cm 1.0cm},clip=true]{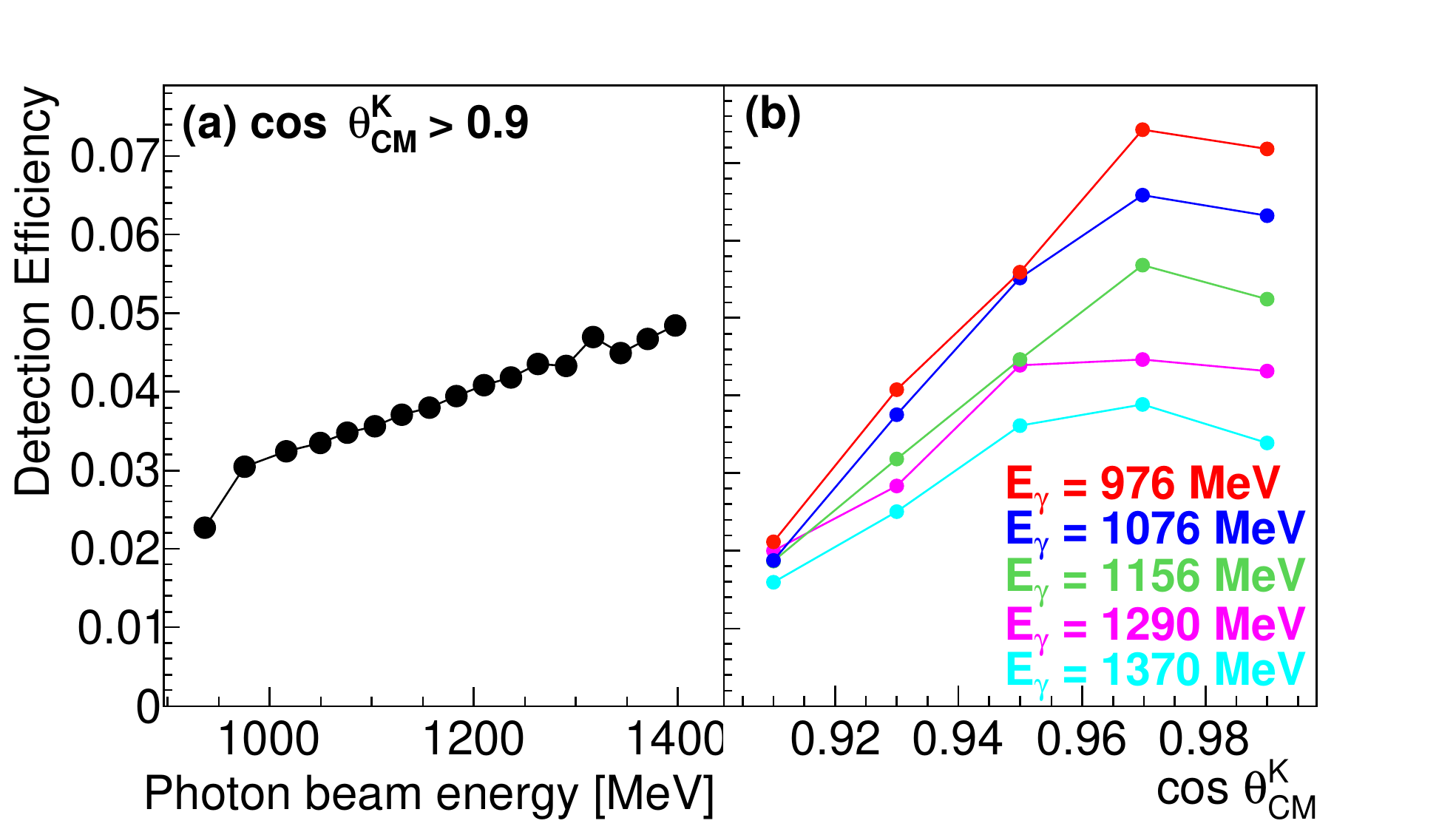} 
	}
	\caption{Detection efficiency for: (a) \kaonangle $> 0.9$ versus photon beam energy and
	(b) versus \kaonangle for selected photon energy intervals labelled inset.  The connecting lines are an aid to guide the eye.}
	\label{fig:deteff}
\end{figure}

 Shown in fig.~\ref{fig:deteff}, the detection efficiency was approximately 2.4\,\% at threshold, rising smoothly to 5\,\% at 1400\,MeV.  The efficiency also increases at more forward angles.  These efficiencies also account for the $\pi^0$ detection, the $\Lambda \rightarrow \pi^0 n$ branching ratio of 36\,\%, and approximately 50\,\% of $K^+$ decaying in-flight.
These three factors alone limit the detection efficiency to 13\,\%.

\subsection{Systematic uncertainties}\label{sec:syserror}

Systematic uncertainties are divided into two components. The \textit{scaling uncertainty},  the sources of which are listed in table~\ref{table:syserror}, is a constant fraction of the measured cross section. The position of the beam when impinging upon the target was the largest source due to the dependence of the measured production angle and forward acceptance. This was determined using simulated data. The absolute photon flux determination is the second largest uncertainty. This was estimated by measuring well known photoproduction cross sections (for example $\gamma p \rightarrow \pi^0p$ and $\eta p$ shown in ref.~\cite{technicalpaper})), and comparing flux measurements using the tagging efficiency calculations from the Flumo and GIM detectors.  Flumo measured the tagging efficiency continuously during the data taking, whereas GIM measured the tagging efficiency every 12 hours at low rates (an extracted electron beam of 40\,pA compared to 1420\,pA).  Despite the different beam conditions, an agreement of the flux normalisation to within 3\,\% was achieved.  The electron beam position upon the diamond radiator was also closely monitored by a continuous study of the coherent edge of the linearly polarised bremsstrahlung photon energy distribution.  
\begin{table}[h]
	\centering
	\begin{tabular}{l c}
		\hline\hline
		Source & \% error \\
		\hline
Beam spot alignment & 4.0 \\
Photon flux & 4.0\\
$K^+$ selection & 2.0 \\
SciFi efficiency & 3.0 \\
Target wall contribution & 2.0 \\
Track time selection & 2.0 \\
Target length & 1.7 \\
ToF wall efficiency & 1.5 \\
MOMO efficiency & 1.0 \\
Drift chamber efficiency & 1.0 \\
Beam energy calibration & 1.0 \\
Modelling of hardware triggers & 1.0 \\
$\pi^0$ identification  & 1.0 \\
Forward track geometric selection & 1.0 \\
\hline
Summed in quadrature & 8.0 \\
\hline\hline
	\end{tabular}
	\caption{Systematic uncertainties contributing to the constant fractional error.}
	\label{table:syserror}
\end{table}

The \textit{fitting uncertainty} from extracting the number of events from the missing mass spectra permits the individual movement of data points. 
 This was estimated from the difference of when including the additional simulated $\Delta^0\pi^+$ events in the background distribution and by also varying the fit range.  An exponential function was fitted to the difference in the cross section to describe the general trend.  The only significant differences were at the four data points at the highest energies where the  signal yield begins to reduce compared to the background and the $K^+$ missing mass distribution becomes broader.  This gave an uncertainty of 0.022 and 0.042\,$\mu$b/sr at centre of mass energies 1831 and 1858\,MeV respectively.  The data stops at 1858\,MeV as this uncertainty becomes very large at higher energies.
 
To check the consistency of the fitting procedure, the data were also binned into
both 0.03 and 0.02 \kaonangle intervals, where the yield was summed and compared to the total over the full 0.1 \kaonangle interval.  This showed good agreement within the  systematic errors.
The same fitting systematic uncertainty was assumed for the data binned in smaller \kaonangle{} intervals, where the reduced statistics prevented an accurate determination.



\section{Results and discussion}\label{sec:results}

All presented data are tabulated in the appendix.  The data extends to a photon beam energy of 1400\,MeV, corresponding to a centre of mass energy of 1858\,MeV.  Above this energy the systematic uncertainty in separating the signal from background begins to increase very quickly. 

\subsection{$\gamma p \rightarrow K^+\Lambda$ differential cross section}

\begin{figure} [htb]
	\centering
	\vspace*{0cm}
	\resizebox{\columnwidth}{!}{%
		\includegraphics{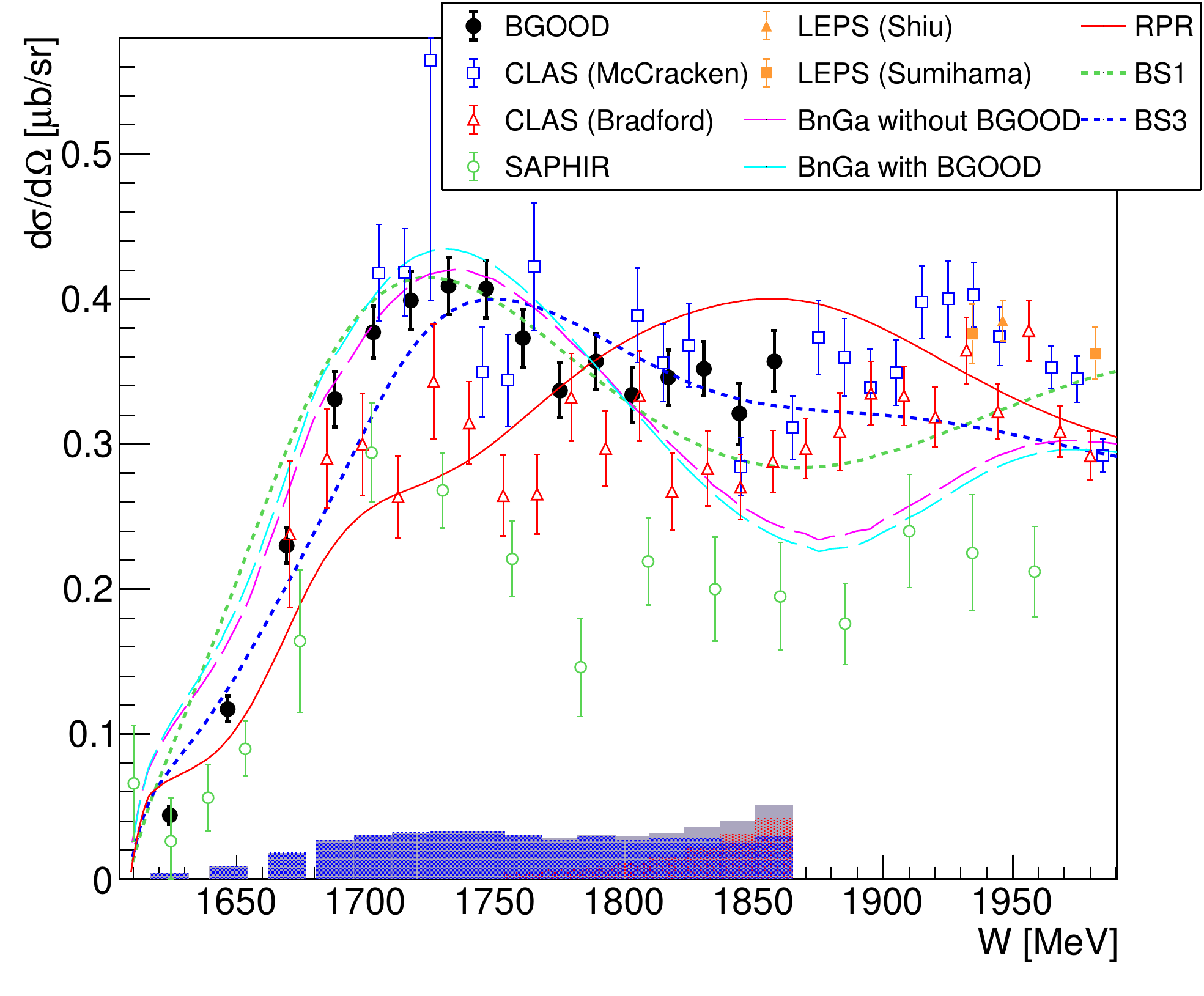} 
	}
	\caption{$\gamma p \rightarrow K^+\Lambda$ differential cross section for \kaonangle $>0.90$	(black filled circles). The systematic uncertainties on the abscissa are in three components: The shaded blue and red bars are the \textit{scaling} and \textit{fitting} uncertainties respectively, described in sec.~\ref{sec:syserror}. The grey bars are the total. Previous data (only including statistical errors) is shown of McCracken \textit{et al.} (CLAS)~\cite{mccracken10} (blue open squares), Bradford \textit{et al.} (CLAS)~\cite{bradford06} (red open triangles), Glander \textit{et al.} (SAPHIR)~\cite{glander04} (green open diamonds), Shiu \textit{et al.} (LEPS)~\cite{shiu18} (orange filled triangle) and Sumihama \textit{et al.}~\cite{sumihama06} (orange filled squares).
		The CLAS data are at the more backward angle of 0.85 $<$\kaonangle{} $<$ 0.95.
		The Regge plus resonant model  \cite{bydovsky19} and isobar models BS1 and BS3  \cite{skoupil16,skoupil18}  of Skoupil and Byd\v{z}ovsk\'{y} are the  solid red, dotted green and  dotted blue lines respectively.
		The Bonn-Gatchina PWA~\cite{muller19} solutions with and without the inclusion of the new data are the dashed cyan and dashed magenta lines respectively.}		
	\label{fig:cstotal}
\end{figure}

The differential cross section for $\cos\theta_{CM}^{K} > 0.9$  is shown in fig.~\ref{fig:cstotal}.
The interval range in $W$ is typically 14\,MeV and determined by the width of the Photon Tagger channels.  This is comparable to the previous data shown from the CLAS collaboration~\cite{bradford06,mccracken10} and half the size of the SAPHIR collaboration data~\cite{glander04}. 
It should be noted that the CLAS data is at the more backward angle of $0.85 < \cos\theta_{CM}^{K} < 0.95$, and the SAPHIR data is the only other dataset at this most forward \kaonangle{} interval.
The statistical error, as a fraction of the measured data, is improved by approximately a factor of two over most of the measured energy range.
 
The available datasets at these forward \kaonangle{} intervals exhibit discrepancies, where the SAPHIR data is consistently lower than the CLAS data, and the two CLAS datasets also deviate from each other.  These new data appear in agreement with the CLAS data of McCracken~\cite{mccracken10}.  The CLAS data of Bradford~\cite{bradford06} appears (by eye) approximately 20\,\% lower for energies below 1850\,MeV  and the SAPHIR data~\cite{glander04} are lower over the full energy range by the order of 30 to 40\,\%.

The isobar models of Skoupil and Byd\v{z}ovsk\'{y} \cite{skoupil16,skoupil18}, BS1 and BS3 (green and blue dotted lines), also plotted in fig.~\ref{fig:cstotal}, show good agreement with the peak structure around $1720$\,MeV.  The data exhibits a flatter structure from 1800 to 1850\,MeV, which the BS3 model appears to reproduce well.  A peak is evident in both the BS1 and BS3 models at this energy but at a more backward angle of \kaonangle\,$\approx 0.4$ which is not covered by this new data.  

The Regge plus resonant (RPR) model of Skoupil and Byd\v{z}ovsk\'{y} \cite{bydovsky19} (red line) fails to reproduce the bump at $1720$\,MeV, where it is considered that the $S_{11}(1650)$ would need to contribute more to describe the data.  This new data with improved statistics will help constrain the RPR model where previously it was fitted to the less precise CLAS and LEPS datasets within this forward region~\cite{dilaborprivate}.
There is an improved agreement  with the RPR model for energies beyond $1800$\,MeV, where the rise is due to the constructive interference of the  $D_{13}$(1700) and $D_{15}$(1675), however the data exhibits a flatter distribution.
Neither resonances are included in the BS1 or BS3 isobar models, which may cause the discrepancies at these energies~\cite{dilaborprivate}.
The flatter distribution of the cross section for energies greater than $1800$\,MeV for this data, the CLAS Bradford data and the LEPS data~\cite{shiu18,sumihama06} is inherent to Regge based models which cannot introduce structure, compared to isobar models.  The RPR model amplitude within this region however is still strongly influenced by the parameters from the $s$ channel contributions, with the Regge region only applicable above 3\,GeV~\cite{dilaborprivate}.

The Bonn-Gatchina BG2019 solution~\cite{muller19}, when fitted simultaneously to both the CLAS data is also shown in fig.~\ref{fig:cstotal} as the magenta line.  There is a reduced $\chi^2$ of 2.99 between the fit and this data.  The fit describes this data well below 1800\,MeV however above this energy the fit reduces in strength and does not reproduce the slight rise of the data points.
A new fit additionally including this data is shown as the cyan line.  The fit optimized all $K^+\Lambda$  and $K^+\Sigma^0$  couplings for the resonant contributions and $t$ and $u$ channel exchange amplitudes with $K^+\Lambda$ and $K^+\Sigma^0$ final states.  Only reactions with two body final states were fitted.  A full parameter optimisation was then made, fitting all reactions from the Bonn-Gatchina PWA database. Finally, all three
body couplings were fixed. The reduced $\chi^2$ between this new fit and the data improved to 2.41. 
The only significant changes occurred in the forward region, with negligible changes to the more backward region covered by the CLAS data.
The inclusion of this data changed contributions
from the non-resonant amplitudes defined by the $K^0$(1430)
and $\Sigma$ exchanges. For the resonant couplings the solution
readjusted the $K\Sigma$ couplings of the highest $P_{11}$ states.
However these readjustments did not significantly change
the absolute values of the couplings calculated as residues
in the pole position, where only relative phases changed by one standard deviation.
The most notable changes were found in the $A^{1/2}$ helicity couplings for the P$_{33}$(1920) and helicity couplings of the P$_{13}$(1900), although in both cases these changed by less than two standard deviations.
The fit was repeated by iteratively adding
resonant contributions with different quantum numbers. 
Only a small improvement of the description could be achieved.
The most notable changes are observed for
resonances with $J^- = 5/2^-$, which provided the best overall improvement, without making any significant change to the more backward CLAS data.

\begin{figure} [htb]
		\includegraphics[width=\columnwidth,trim={0cm 2.5cm 0cm 0cm},clip=true]{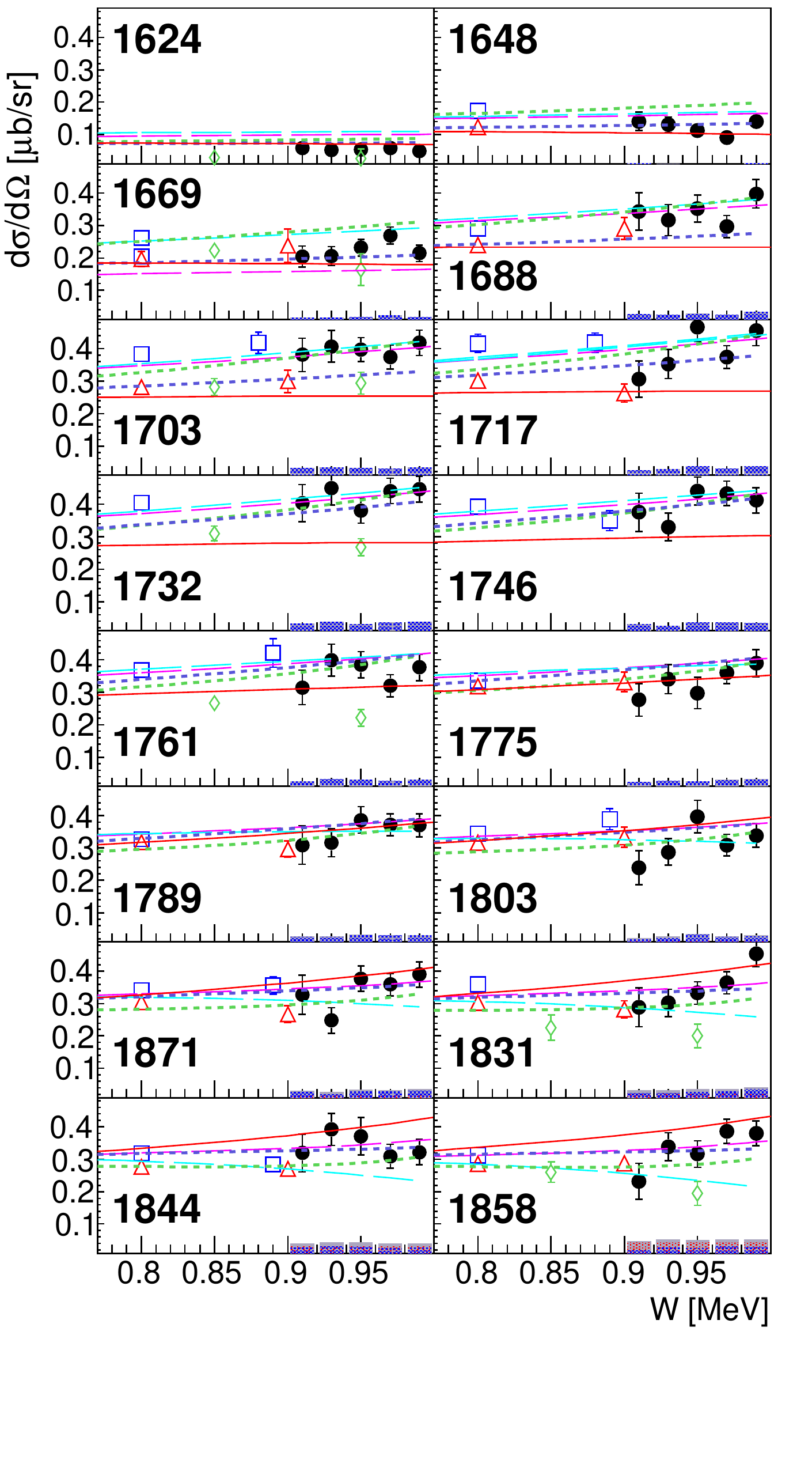} 
	\caption{$\gamma p \rightarrow K^+\Lambda$ differential cross section versus \kaonangle{} for each centre of mass energy, $W$ labelled inset in MeV.  Filled black circles are these data binned into 0.02 \kaonangle{} intervals, and other data points and model fits are the same as described in fig.~\ref{fig:cstotal}.
}
	\label{fig:csvsangle}
\end{figure}

\begin{figure} [htb]
	\centering
	\vspace*{0cm}
	\resizebox{\columnwidth}{!}{%
		\includegraphics[width=\columnwidth,trim={0cm 2cm 0.5cm 0.6cm},clip=true]{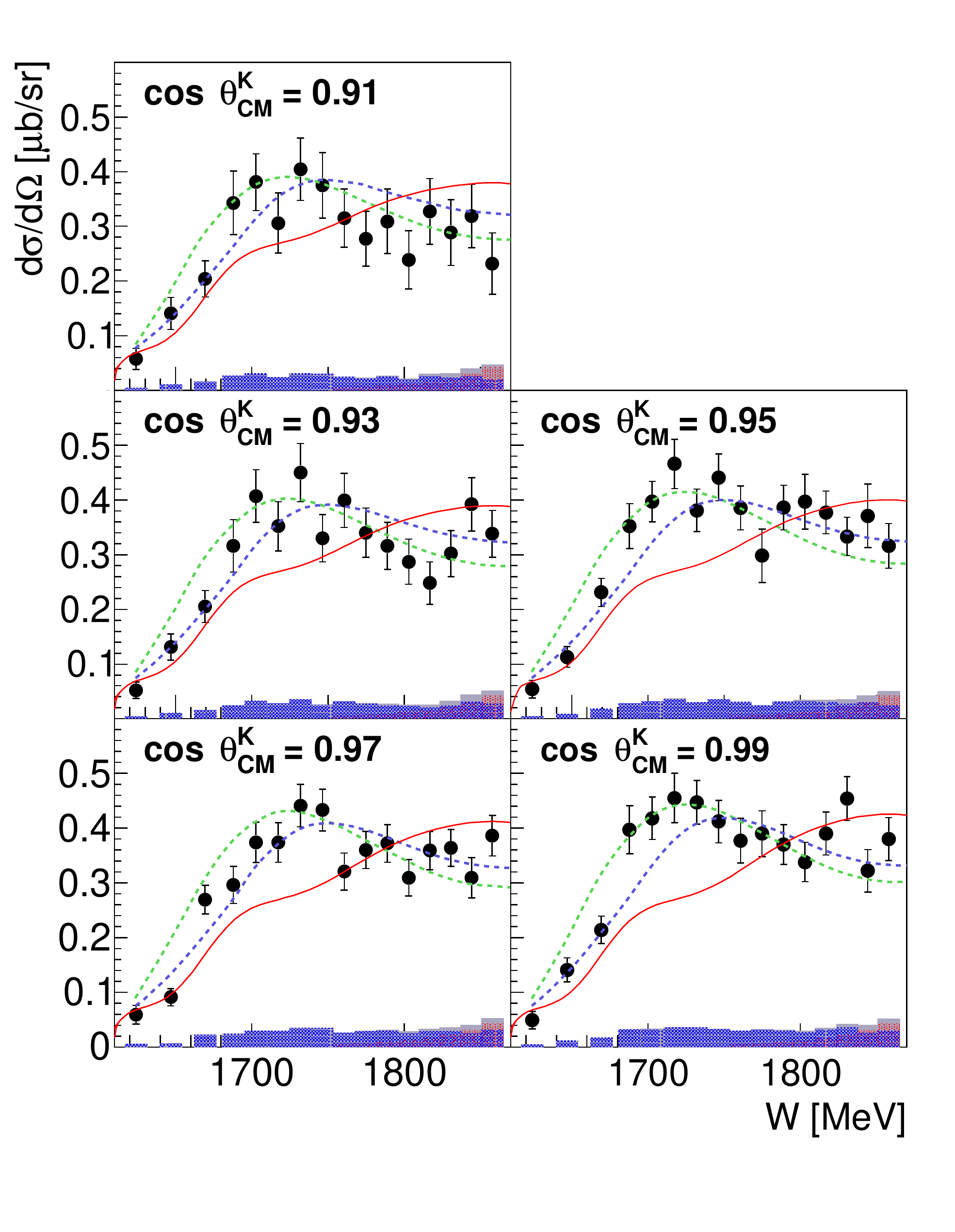} 
	}
	\caption{$\gamma p \rightarrow K^+\Lambda$ differential cross section for intervals of 0.02 in \kaonangle (filled black circles).  Other data points and model fits are the same as described in fig.~\ref{fig:cstotal}.
	}
	\label{fig:csvsenergyfine}
\end{figure}

 Figures \ref{fig:csvsangle} and \ref{fig:csvsenergyfine} show the differential cross section in 0.02 \kaonangle{} intervals versus \kaonangle{} and $W$
respectively.
Near threshold, the distribution is flat, suggesting $s$-channel dominating components of the reaction mechanism. As $W$ increases the cross section becomes more forward peaked consistent with increasing $t$-channel $K$ and $K^*$ exchange processes.
In fig.~\ref{fig:csvsenergyfine},  the peak at $1720$\,MeV remains approximately constant in strength over the \kaonangle{} range.

\begin{figure} [h]
	\centering
	\vspace*{0cm}
	\resizebox{0.5\textwidth}{!}{%
					\includegraphics[width=\columnwidth,trim={0cm 2.5cm 0cm 3.5cm},clip=true]{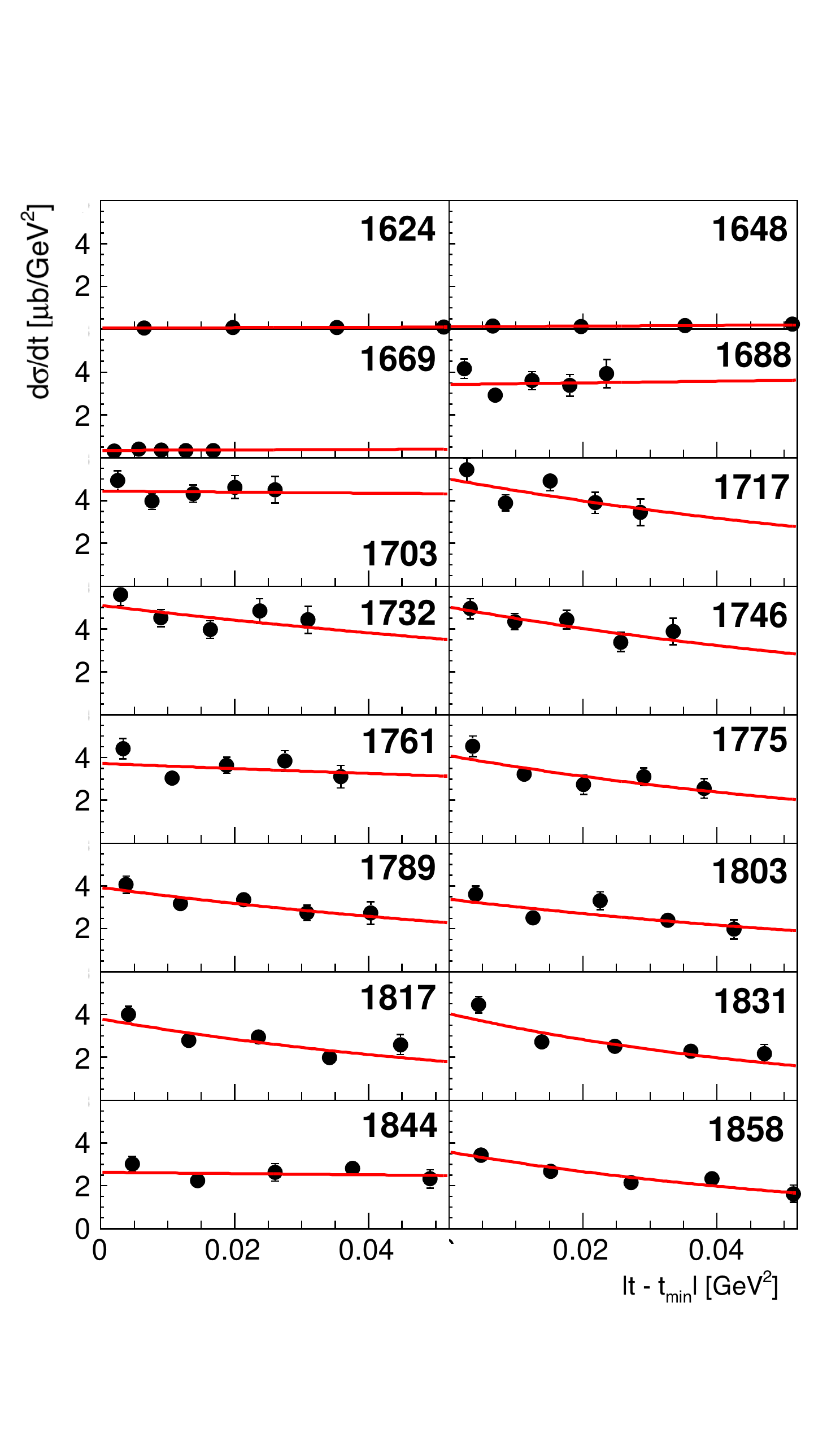} 
	}
	\caption{d$\sigma$/d$t$ versus $|t - t_\mathrm{min}|$ for intervals of centre of mass energy, $W$, labelled inset in MeV.  Only the statistical error is shown and included in the fit.  The red line is eq.~1 fitted to the data.
	}
	\label{fig:fittingslope}
\end{figure}

The data binned finely into 0.02 \kaonangle~intervals was used to determine the differential cross section with respect to the Mandelstam variable, $t = (p_\gamma - p_K)^2$, where $p_\gamma$ and $p_K$ are the four-momenta of the photon beam and $K^+$ respectively.  To account for the distribution of $t$ within each two dimensional $W$ and \kaonangle{} interval, a generated distribution assumed the differential cross section of the McCracken CLAS data~\cite{mccracken10}.  For each interval of the BGOOD data in $W$ and \kaonangle, the mean average value of $t$ was used as the central value, and the width was determined as $\sqrt{12}$\,RMS.
 The BGOOD differential cross section data with respect to $t$ is shown for each $W$ interval  in fig.~\ref{fig:fittingslope}.  The function in eq.~1  was fitted to the data to interpolate the cross section to the minimum value of $t$ achievable for the given $W$ interval, $t_\mathrm{min}$ (occurring at \kaonangle $= 1$), and to extract the slope parameter, $S$.

\begin{equation}
\frac{\mathrm{d}\sigma}{\mathrm{d}t} = \frac{\mathrm{d}\sigma}{\mathrm{d}t}\Big|_{t=t_\mathrm{min}}e^{S|t-t_\mathrm{min}|}
\end{equation}\label{eq:fitfunction}

\begin{figure} [h]
	\centering
		\includegraphics[width=\columnwidth,trim={0cm 1.7cm 0cm 0.5cm},clip=true]{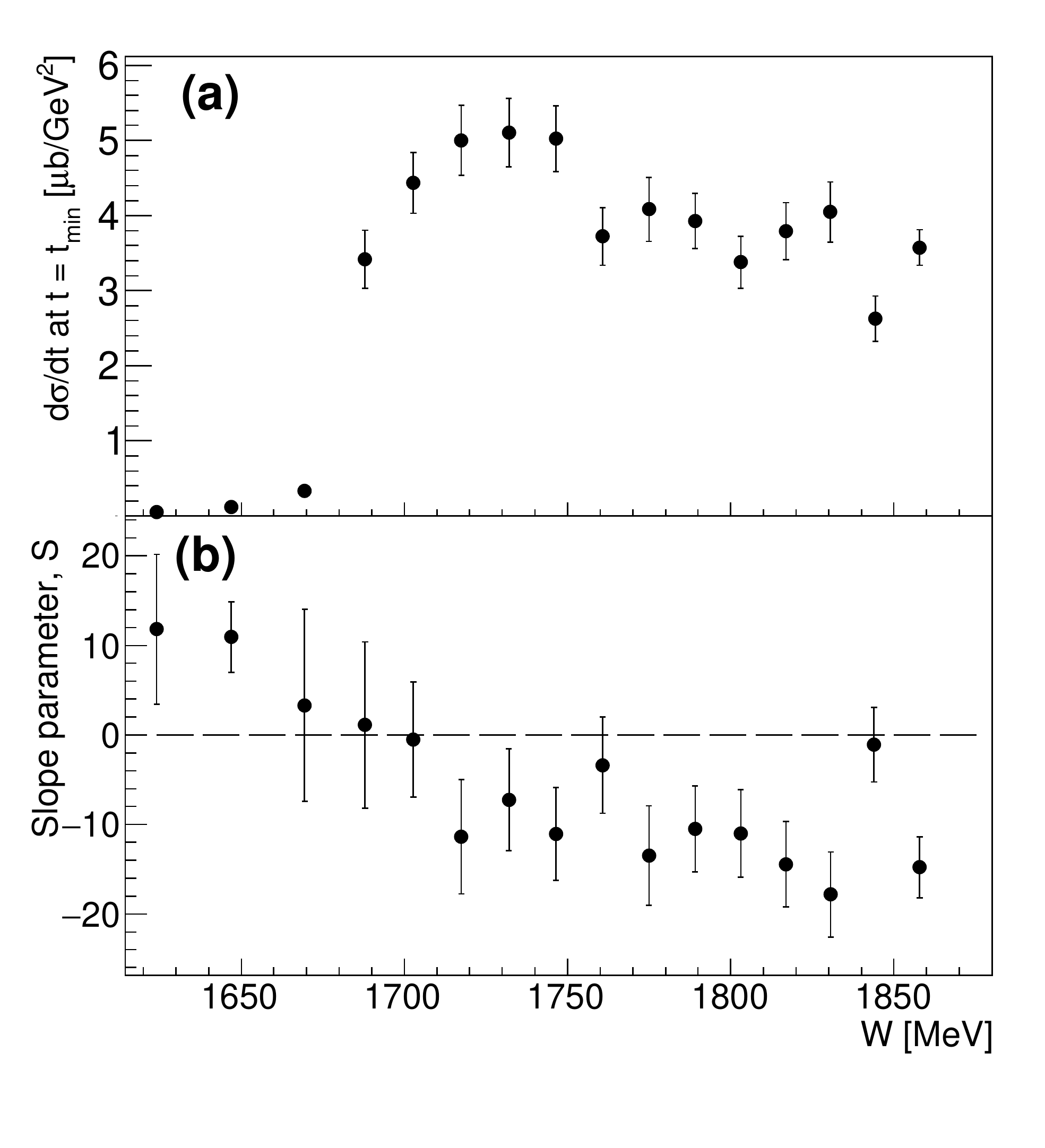}
	\caption{(a) $K^+\Lambda$ differential cross section, d$\sigma$/d$t$ extrapolated to $t_\mathrm{min}$ versus $W$. (b) The slope parameter $S$ versus $W$.
	}
	\label{fig:cstmin}
\end{figure}

Fig.~\ref{fig:cstmin} shows the differential cross section at $t_\mathrm{min}$
 and the slope parameter $S$ versus $W$.   The shape of the cross section is similar to the most forward \kaonangle{} interval, with a dominant peak at 1720\,MeV.  For the first 100\,MeV above threshold, $S$ remains positive.
 At higher energies, $S$ becomes increasingly negative, indicating the onset of $t$-channel $K$ exchange dominating the reaction mechanism.

\subsection{$\gamma p \rightarrow K^+\Lambda$ recoil polarisation}

\begin{figure} [h]
 	\centering
 	\vspace*{0cm}
 	\resizebox{0.9\columnwidth}{!}{%
 		\centering
 		\includegraphics{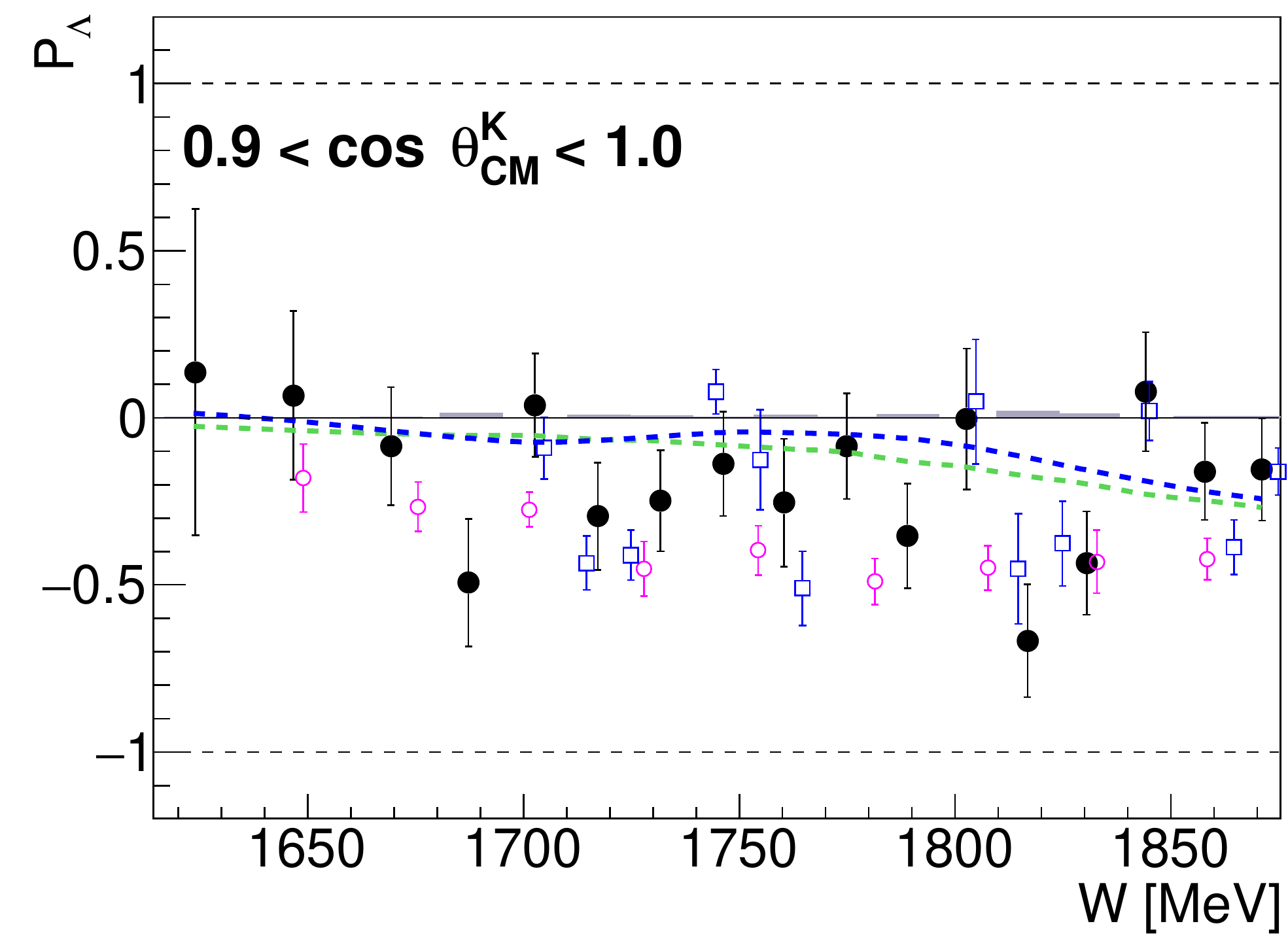} 
 	}
 	\caption{Recoil polarisation, $P_\Lambda$ for 0.9 $<$\kaonangle$ < 1.0$ (black circles).  Previous data (only including statistical errors) of McCracken \textit{et al.} (CLAS)~\cite{mccracken10} for $<0.85$ \kaonangle$<0.95$ and Lleres \textit{et al.} (GRAAL)~\cite{lleres07} for approximately $0.77 <$ \kaonangle$ < 0.94$ shown as blue open squares and magenta open circles respectively.  The two isobar models,  BS1 and BS3 of Skoupil and Byd\v{z}ovsk\'{y}~\cite{skoupil16,skoupil18} are the dotted green and blue lines respectively.}
 	\label{fig:recpol}
 \end{figure}

The weak decay of the $\Lambda$ allows access to the recoil polarisation via the decay distribution. The $\pi^0$ four-momentum from $\Lambda \rightarrow \pi^0 n$  was boosted into the $\Lambda$ rest frame and the $\pi^0$ direction relative to the reaction plane was determined (denoted $N_{\uparrow/\downarrow}$).   The recoil polarisation was measured according to eq.~2.  The $\Lambda$ decay parameter used,  $\alpha = 0.642 \pm 0.04$~\cite{pdg2018} is the average value cited by the Particle Data Group prior to 2019\footnote{This older value of $\alpha$ was chosen for consistency as the isobar models of Skoupil and Byd\v{z}ovsk\'{y}~\cite{skoupil16,skoupil18} shown in fig.~\ref{fig:recpol} are fitted to a combination of data which used this.  The value since 2019, $\alpha = 0.732 \pm 0.014$~\cite{pdg20}
would reduce all data points shown and associated errors by a factor of 0.877.}.  
 
 \begin{equation}
 P_\Lambda = \frac{2}{\alpha}\frac{N_\uparrow - N_\downarrow}{N_\uparrow + N_\downarrow}
 \end{equation}\label{eq:recpol}
 
 Simulated data were used to determine the success rate of correctly determining $N_{\uparrow/\downarrow}$ per event to measure dilution effects which may have occurred due to limited azimuthal angular resolution at forward angles. A small correction as a function of $E_\gamma$ was determined.  This was 5\,\% and 7\,\% at $E_\gamma = 914$\,MeV (threshold) and 1400\,MeV respectively.
 
 
The recoil polarisation data is shown in fig.~\ref{fig:recpol}.  The systematic uncertainties shown in table~\ref{table:syserror} and the fitting uncertainty mostly cancel out. The remaining dominating uncertainty is the accuracy of $\alpha$ of 6.2\,\%.
 
This is the first data for $P_\Lambda$ in this most forward \kaonangle{} interval (the previous data shown are at more backward angles described in the figure caption).
$P_\Lambda$ is consistent with zero at threshold and at higher energies becomes negative, consistent with the isobar models, BS1 and BS3~\cite{skoupil16,skoupil18}.
The Bonn-Gatchina BG2019 solution prior to including this data gives a $\chi^2$ of 0.98 for the recoil asymmetry.  When refitting using the new data as described above, $\chi^2$ changes to 0.95.
 

\section{Conclusions}\label{sec:conclusions}

Differential cross sections for $\gamma p \rightarrow K^+\Lambda$ for \kaonangle{} $> 0.9$ have been measured with high polar angle resolution from threshold to $W = 1870$\,MeV.  
A consistency is observed between this data and the CLAS data of McCracken \textit{et al.}~\cite{mccracken10}, which is also supported by a dedicated Bonn Gatchina PWA analysis.  
The high statistics provide constraints in determining dominating $t$-channel $K$ and $K^*$ exchange at forward angles and low momentum transfer, and the \kaonangle{} resolution renders the data particularly sensitive to intermediate high-spin states.  Additionally, the recoil polarisation data for $K^+\Lambda$ is the first dataset at this most forward \kaonangle{} interval.  

 
 \section*{Acknowledgements}
 
  We thank the staff and shift-students of the ELSA accelerator for providing an excellent beam.  
 
 We thank Dalibor Skoupil and Petr Byd\v{z}ovsk\'{y} for insightful input and comparison of the data to their isobar and RPR models
   and Eberhard Klempt for help with the Bonn-Gatchina PWA.
 
  
 This work is supported by SFB/TR-16, DFG project numbers 388979758 and 405882627, the RSF grant numbers 19-42-04132 and 19-12-04132, the Third Scientific Committee of the INFN and the European Union’s Horizon 2020 
 research and innovation programme under grant agreement number 824093.
 P.L. Cole gratefully acknowledges the support from both the U.S. National Science Foundation (NSF-PHY-1307340, NSF-PHY-1615146, and NSF-PHY-2012826) and the Fulbright U.S. Scholar Program (2014/2015).
 
\bibliographystyle{unsrt}

\setcounter{tocdepth}{3}
\newpage
\section*{Appendix: Tabulated data}

\begin{table}[h]
	\centering
	\begin{tabular}{ c c c c c c c}
		\hline\hline
			&&&&&&\\
		\multicolumn{7}{l}{$0.90<$\kaonangle$<1.00$} \\
		\hline
		$W$ &  $\Delta W$ & d$\sigma$/d$\Omega$ & $\delta_\mathrm{stat}$ &$\delta_\mathrm{sys}$&  $\delta_\mathrm{scaling}$ &  $\delta_\mathrm{fitting}$ \\ 	 

		MeV & MeV &$\mu$b/sr & $\mu$b/sr & $\mu$b/sr & $\mu$b/sr & $\mu$b/sr\\
		\hline
		
		1624.1	&	23.0	&	0.044	&	0.006	&	0.004	&	0.004	&	0.000	\\
		1647.0	&	22.7	&	0.117	&	0.009	&	0.009	&	0.009	&	0.000	\\
		1669.4	&	22.3	&	0.230	&	0.012	&	0.018	&	0.018	&	0.000	\\
		1688.0	&	14.9	&	0.331	&	0.019	&	0.027	&	0.027	&	0.000	\\
		1702.8	&	14.7	&	0.377	&	0.018	&	0.030	&	0.030	&	0.000	\\
		1717.4	&	14.6	&	0.399	&	0.020	&	0.032	&	0.032	&	0.000	\\
		1732.0	&	14.5	&	0.409	&	0.020	&	0.033	&	0.033	&	0.000	\\
		1746.4	&	14.4	&	0.407	&	0.020	&	0.033	&	0.033	&	0.000	\\
		1760.8	&	14.3	&	0.373	&	0.020	&	0.030	&	0.030	&	0.005	\\
		1775.0	&	14.1	&	0.337	&	0.019	&	0.028	&	0.027	&	0.007	\\
		1789.0	&	14.0	&	0.357	&	0.019	&	0.030	&	0.029	&	0.009	\\
		1803.0	&	13.9	&	0.334	&	0.019	&	0.029	&	0.027	&	0.011	\\
		1816.9	&	13.8	&	0.346	&	0.019	&	0.032	&	0.028	&	0.016	\\
		1830.7	&	13.7	&	0.352	&	0.019	&	0.036	&	0.028	&	0.022	\\
		1844.3	&	13.6	&	0.321	&	0.021	&	0.040	&	0.026	&	0.031	\\
		1857.9	&	13.5	&	0.357	&	0.021	&	0.051	&	0.029	&	0.042	\\

		\hline\hline
\end{tabular}\caption{$\gamma p \rightarrow K^+\Lambda$ differential cross section data (d$\sigma$/d$\Omega$) for $0.90<$\kaonangle$<1.00$.  The median and width of each centre of mass interval are labelled $W$ and $\Delta W$respectively.  The statistical, systematic, and the two components of the systematic error (scaling and fitting) are labelled $\delta_\mathrm{stat}$, $\delta_\mathrm{sys}$, $\delta_\mathrm{scaling}$ and $\delta_\mathrm{fitting}$ respectively.}\label{table:cs1}
\end{table}

\begin{table}[h]
	\centering
	\begin{tabular}{ c c c c c c c}
		\hline\hline
		&&&&&&\\
		\multicolumn{7}{l}{$0.90<$\kaonangle$<0.92$} \\
		\hline
		$W$ &  $\Delta W$ & d$\sigma$/d$\Omega$ & $\delta_\mathrm{stat}$ &$\delta_\mathrm{sys}$&  $\delta_\mathrm{scaling}$ &  $\delta_\mathrm{fitting}$ \\ 	 
		
		MeV & MeV &$\mu$b/sr & $\mu$b/sr & $\mu$b/sr & $\mu$b/sr & $\mu$b/sr\\
		\hline
1624.1	&	23.0	&	0.058	&	0.019	&	0.005	&	0.005	&	0.000	\\
1647.0	&	22.7	&	0.141	&	0.029	&	0.011	&	0.011	&	0.000	\\
1669.4	&	22.3	&	0.204	&	0.033	&	0.016	&	0.016	&	0.000	\\
1688.0	&	14.9	&	0.343	&	0.058	&	0.027	&	0.027	&	0.000	\\
1702.8	&	14.7	&	0.381	&	0.052	&	0.031	&	0.031	&	0.000	\\
1717.4	&	14.6	&	0.306	&	0.055	&	0.024	&	0.024	&	0.000	\\
1732.0	&	14.5	&	0.404	&	0.057	&	0.032	&	0.032	&	0.000	\\
1746.4	&	14.4	&	0.375	&	0.060	&	0.030	&	0.030	&	0.000	\\
1760.8	&	14.3	&	0.315	&	0.053	&	0.025	&	0.025	&	0.005	\\
1775.0	&	14.1	&	0.277	&	0.050	&	0.023	&	0.022	&	0.007	\\
1789.0	&	14.0	&	0.309	&	0.059	&	0.027	&	0.025	&	0.009	\\
1803.0	&	13.9	&	0.239	&	0.053	&	0.022	&	0.019	&	0.012	\\
1816.9	&	13.8	&	0.327	&	0.060	&	0.031	&	0.026	&	0.016	\\
1830.7	&	13.7	&	0.289	&	0.060	&	0.032	&	0.023	&	0.022	\\
1844.3	&	13.6	&	0.319	&	0.058	&	0.040	&	0.026	&	0.031	\\
1857.9	&	13.5	&	0.232	&	0.056	&	0.047	&	0.019	&	0.043	\\
		\hline\hline
	\end{tabular}\caption{$\gamma p \rightarrow K^+\Lambda$ differential cross section data (d$\sigma$/d$\Omega$) for $0.90<$\kaonangle$<0.92$.  The notation is the same as in table~\ref{table:cs1}.}
\end{table}

\begin{table}[h]
	\centering
	\begin{tabular}{ c c c c c c c}
		\hline\hline
		&&&&&&\\
		\multicolumn{7}{l}{$0.92<$\kaonangle$<0.94$} \\
		\hline
		$W$ &  $\Delta W$ & d$\sigma$/d$\Omega$ & $\delta_\mathrm{stat}$ &$\delta_\mathrm{sys}$&  $\delta_\mathrm{scaling}$ &  $\delta_\mathrm{fitting}$ \\ 	 
		
		MeV & MeV &$\mu$b/sr & $\mu$b/sr & $\mu$b/sr & $\mu$b/sr & $\mu$b/sr\\
		\hline
1624.1	&	23.0	&	0.052	&	0.015	&	0.004	&	0.004	&	0.000	\\
1647.0	&	22.7	&	0.131	&	0.024	&	0.010	&	0.010	&	0.000	\\
1669.4	&	22.3	&	0.205	&	0.029	&	0.016	&	0.016	&	0.000	\\
1688.0	&	14.9	&	0.316	&	0.048	&	0.025	&	0.025	&	0.000	\\
1702.8	&	14.7	&	0.407	&	0.048	&	0.033	&	0.033	&	0.000	\\
1717.4	&	14.6	&	0.352	&	0.045	&	0.028	&	0.028	&	0.000	\\
1732.0	&	14.5	&	0.450	&	0.053	&	0.036	&	0.036	&	0.000	\\
1746.4	&	14.4	&	0.330	&	0.043	&	0.026	&	0.026	&	0.000	\\
1760.8	&	14.3	&	0.399	&	0.049	&	0.032	&	0.032	&	0.005	\\
1775.0	&	14.1	&	0.340	&	0.045	&	0.028	&	0.027	&	0.007	\\
1789.0	&	14.0	&	0.316	&	0.043	&	0.027	&	0.025	&	0.009	\\
1803.0	&	13.9	&	0.287	&	0.041	&	0.026	&	0.023	&	0.012	\\
1816.9	&	13.8	&	0.248	&	0.039	&	0.026	&	0.020	&	0.016	\\
1830.7	&	13.7	&	0.302	&	0.042	&	0.033	&	0.024	&	0.022	\\
1844.3	&	13.6	&	0.392	&	0.049	&	0.044	&	0.031	&	0.031	\\
1857.9	&	13.5	&	0.338	&	0.043	&	0.051	&	0.027	&	0.043	\\
		\hline\hline
	\end{tabular}\caption{$\gamma p \rightarrow K^+\Lambda$ differential cross section data (d$\sigma$/d$\Omega$) for $0.92<$\kaonangle$<0.94$.  The notation is the same as in table~\ref{table:cs1}.}
\end{table}

\begin{table}[h]
	\centering
	\begin{tabular}{ c c c c c c c}
		\hline\hline
		&&&&&&\\
		\multicolumn{7}{l}{$0.94<$\kaonangle$<0.96$} \\
		\hline
		$W$ &  $\Delta W$ & d$\sigma$/d$\Omega$ & $\delta_\mathrm{stat}$ &$\delta_\mathrm{sys}$&  $\delta_\mathrm{scaling}$ &  $\delta_\mathrm{fitting}$ \\ 	 
		
		MeV & MeV &$\mu$b/sr & $\mu$b/sr & $\mu$b/sr & $\mu$b/sr & $\mu$b/sr\\
		\hline
1624.1	&	23.0	&	0.054	&	0.016	&	0.004	&	0.004	&	0.000	\\
1647.0	&	22.7	&	0.113	&	0.019	&	0.009	&	0.009	&	0.000	\\
1669.4	&	22.3	&	0.231	&	0.026	&	0.018	&	0.018	&	0.000	\\
1688.0	&	14.9	&	0.352	&	0.041	&	0.028	&	0.028	&	0.000	\\
1702.8	&	14.7	&	0.397	&	0.037	&	0.032	&	0.032	&	0.000	\\
1717.4	&	14.6	&	0.466	&	0.045	&	0.037	&	0.037	&	0.000	\\
1732.0	&	14.5	&	0.381	&	0.039	&	0.030	&	0.030	&	0.000	\\
1746.4	&	14.4	&	0.441	&	0.043	&	0.035	&	0.035	&	0.000	\\
1760.8	&	14.3	&	0.385	&	0.040	&	0.031	&	0.031	&	0.005	\\
1775.0	&	14.1	&	0.298	&	0.049	&	0.025	&	0.024	&	0.007	\\
1789.0	&	14.0	&	0.386	&	0.041	&	0.032	&	0.031	&	0.009	\\
1803.0	&	13.9	&	0.397	&	0.050	&	0.034	&	0.032	&	0.012	\\
1816.9	&	13.8	&	0.377	&	0.039	&	0.034	&	0.030	&	0.016	\\
1830.7	&	13.7	&	0.333	&	0.035	&	0.035	&	0.027	&	0.022	\\
1844.3	&	13.6	&	0.371	&	0.058	&	0.043	&	0.030	&	0.031	\\
1857.9	&	13.5	&	0.316	&	0.041	&	0.050	&	0.025	&	0.043	\\
		\hline\hline
	\end{tabular}\caption{$\gamma p \rightarrow K^+\Lambda$ differential cross section data (d$\sigma$/d$\Omega$) for $0.94<$\kaonangle$<0.96$.  The notation is the same as in table~\ref{table:cs1}.}
\end{table}

\begin{table}[h]
	\centering
	\begin{tabular}{ c c c c c c c}
		\hline\hline
		&&&&&&\\
		\multicolumn{7}{l}{$0.96<$\kaonangle$<0.98$} \\
		\hline
		$W$ &  $\Delta W$ & d$\sigma$/d$\Omega$ & $\delta_\mathrm{stat}$ &$\delta_\mathrm{sys}$&  $\delta_\mathrm{scaling}$ &  $\delta_\mathrm{fitting}$ \\ 	 
		
		MeV & MeV &$\mu$b/sr & $\mu$b/sr & $\mu$b/sr & $\mu$b/sr & $\mu$b/sr\\
		\hline
1624.1	&	23.0	&	0.059	&	0.017	&	0.005	&	0.005	&	0.000	\\
1647.0	&	22.7	&	0.091	&	0.016	&	0.007	&	0.007	&	0.000	\\
1669.4	&	22.3	&	0.269	&	0.026	&	0.022	&	0.022	&	0.000	\\
1688.0	&	14.9	&	0.296	&	0.034	&	0.024	&	0.024	&	0.000	\\
1702.8	&	14.7	&	0.374	&	0.037	&	0.030	&	0.030	&	0.000	\\
1717.4	&	14.6	&	0.374	&	0.036	&	0.030	&	0.030	&	0.000	\\
1732.0	&	14.5	&	0.441	&	0.039	&	0.035	&	0.035	&	0.000	\\
1746.4	&	14.4	&	0.433	&	0.038	&	0.035	&	0.035	&	0.000	\\
1760.8	&	14.3	&	0.321	&	0.034	&	0.026	&	0.026	&	0.005	\\
1775.0	&	14.1	&	0.360	&	0.034	&	0.030	&	0.029	&	0.007	\\
1789.0	&	14.0	&	0.372	&	0.034	&	0.031	&	0.030	&	0.009	\\
1803.0	&	13.9	&	0.309	&	0.033	&	0.028	&	0.025	&	0.012	\\
1816.9	&	13.8	&	0.359	&	0.035	&	0.033	&	0.029	&	0.016	\\
1830.7	&	13.7	&	0.364	&	0.034	&	0.036	&	0.029	&	0.022	\\
1844.3	&	13.6	&	0.309	&	0.037	&	0.040	&	0.025	&	0.031	\\
1857.9	&	13.5	&	0.386	&	0.037	&	0.053	&	0.031	&	0.043	\\
		\hline\hline
	\end{tabular}\caption{$\gamma p \rightarrow K^+\Lambda$ differential cross section data (d$\sigma$/d$\Omega$) for $0.96<$\kaonangle$<0.98$.  The notation is the same as in table~\ref{table:cs1}.}
\end{table}

\begin{table}[h]
	\centering
	\begin{tabular}{ c c c c c c c}
		\hline\hline
		&&&&&&\\
		\multicolumn{7}{l}{$0.98<$\kaonangle$<1.00$} \\
		\hline
		$W$ &  $\Delta W$ & d$\sigma$/d$\Omega$ & $\delta_\mathrm{stat}$ &$\delta_\mathrm{sys}$&  $\delta_\mathrm{scaling}$ &  $\delta_\mathrm{fitting}$ \\ 	 
		
		MeV & MeV &$\mu$b/sr & $\mu$b/sr & $\mu$b/sr & $\mu$b/sr & $\mu$b/sr\\
		\hline
1624.1	&	23.0	&	0.049	&	0.016	&	0.004	&	0.004	&	0.000	\\
1647.0	&	22.7	&	0.141	&	0.022	&	0.011	&	0.011	&	0.000	\\
1669.4	&	22.3	&	0.214	&	0.025	&	0.017	&	0.017	&	0.000	\\
1688.0	&	14.9	&	0.397	&	0.044	&	0.032	&	0.032	&	0.000	\\
1702.8	&	14.7	&	0.418	&	0.039	&	0.033	&	0.033	&	0.000	\\
1717.4	&	14.6	&	0.455	&	0.045	&	0.036	&	0.036	&	0.000	\\
1732.0	&	14.5	&	0.447	&	0.040	&	0.036	&	0.036	&	0.000	\\
1746.4	&	14.4	&	0.412	&	0.039	&	0.033	&	0.033	&	0.000	\\
1760.8	&	14.3	&	0.377	&	0.041	&	0.030	&	0.030	&	0.005	\\
1775.0	&	14.1	&	0.390	&	0.042	&	0.032	&	0.031	&	0.007	\\
1789.0	&	14.0	&	0.370	&	0.036	&	0.031	&	0.030	&	0.009	\\
1803.0	&	13.9	&	0.338	&	0.036	&	0.030	&	0.027	&	0.012	\\
1816.9	&	13.8	&	0.390	&	0.039	&	0.035	&	0.031	&	0.016	\\
1830.7	&	13.7	&	0.454	&	0.040	&	0.042	&	0.036	&	0.022	\\
1844.3	&	13.6	&	0.322	&	0.039	&	0.040	&	0.026	&	0.031	\\
1857.9	&	13.5	&	0.380	&	0.039	&	0.052	&	0.030	&	0.043	\\
		\hline\hline
	\end{tabular}\caption{$\gamma p \rightarrow K^+\Lambda$ differential cross section data (d$\sigma$/d$\Omega$) for $0.98<$\kaonangle$<1.00$.  The notation is the same as in table~\ref{table:cs1}.}
\end{table}

\begin{table}[h]
	\centering
	\begin{tabular}{ c c c c c}
		\hline\hline
		&&&&\\
		\multicolumn{5}{l}{$0.90<$\kaonangle$<1.00$} \\
		\hline
		$W$ &  $\Delta W$ & $P_\Lambda$ & $\delta_\mathrm{stat}$ &$\delta_\mathrm{sys}$ \\ 	 
		
		MeV & MeV & &  & \\
		\hline
1624.1	& 23.0    &		0.131	&	0.488	&	0.004	\\
1647.0	& 22.7    &		0.061	&	0.252	&	0.002	\\
1669.4	& 22.3    &		-0.086	&	0.176	&	0.003	\\
1688.0	& 14.9    &		-0.499	&	0.191	&	0.015	\\
1702.8	& 14.7    &		0.034	&	0.155	&	0.001	\\
1717.4	& 14.6    &		-0.299	&	0.161	&	0.009	\\
1732.0	& 14.5    &		-0.249	&	0.152	&	0.007	\\
1746.4	& 14.4    &		-0.143	&	0.156	&	0.004	\\
1760.8	& 14.3    &		-0.255	&	0.191	&	0.008	\\
1775.0	& 14.1    &		-0.089	&	0.158	&	0.003	\\
1789.0	& 14.0    &		-0.355	&	0.156	&	0.011	\\
1803.0	& 13.9    &		-0.004	&	0.210	&	0.000	\\
1816.9	& 13.8    &		-0.672	&	0.169	&	0.020	\\
1830.7	& 13.7    &		-0.437	&	0.155	&	0.013	\\
1844.3	& 13.6    &		0.076	&	0.178	&	0.002	\\
1857.9	& 13.5    &		-0.162	&	0.145	&	0.005	\\
		
		\hline\hline
	\end{tabular}\caption{$\gamma p \rightarrow K^+\Lambda$ recoil polarisation ($P_\Lambda$) for $0.90<$\kaonangle$<1.00$.  The notation is the same as in table~\ref{table:cs1}, except that only the total systematic error is given.  The $\Lambda$ decay parameter used,  $\alpha = 0.642 \pm 0.04$~\cite{pdg2018} is the average value cited by the Particle Data Group prior to 2019.}
\end{table}

\end{document}